\documentclass[fleqn,usenatbib]{mnras}

\usepackage[T1]{fontenc}
\usepackage{graphicx}
\usepackage{xcolor}
\usepackage{color}
\usepackage{float}
\usepackage{hyperref}
\usepackage{float}
\usepackage{array}
\usepackage{graphicx} 
\usepackage{multirow}
\usepackage{hhline,booktabs}
\usepackage{makecell}
\usepackage{amsmath,amssymb}
\usepackage{newtxtext,newtxmath}
\usepackage{comment}
\hypersetup{colorlinks=true,linkcolor=[rgb]{1.,0.2,0.2},citecolor=[rgb]{0.1,0.4,1.},filecolor=[rgb]{0.7,0.2,0.2},urlcolor=[rgb]{0.0,0.2,1.}}
\makeatletter
\@fleqnfalse
\@mathmargin\@centering
\makeatother

\def\INSPIRE{\mbox{{\tt INSPIRE}}}
\def\EINSPIRE{\mbox{{\tt E-INSPIRE}}}
\newcommand{\Reff}{$\mathrm{R}_{\mathrm{e}\,}$}

\newcommand{\kms}{km s$^{-1}$}
\newcommand{\Msun}{M$_{\odot}\,$}
\usepackage[normalem]{ulem}
\newcommand{\ppxf}{\textsc{pPXF}}

\definecolor{darkgreen}{rgb}{0.09, 0.45, 0.27}
\definecolor{amber}{rgb}{1.0, 0.49, 0.0}

\defcitealias{Spiniello24}{S24}
\defcitealias{Mills25}{\EINSPIRE\ I}

\title[Finding relics in wide-sky surveys]{\centering {\tt E-INSPIRE} - II. Finding relics from wide-sky multi-band surveys: \\ A proof-of-concept machine learning regression algorithm}
\author[C.~Rosen et al.]{\noindent
Charles Rosen$^{1}$, Chiara Spiniello$^{2, 1, 3}$, John Mills$^{4}$\thanks{E-mail: John.N.S.Mills@warwick.ac.uk}, Alexey Sergeyev$^{5,6,7}$, Vladyslav Khramtsov$^{8}$, \and Anna Ferr\'e-Mateu$^{9,10}$, Johanna Hartke$^{11,12,13}$,  Michalina Maksymowicz-Maciata$^{14}$, \and Malgorzata Siudek$^{9,10}$, and Crescenzo Tortora$^{3}$
\\ 
$^{1}$Sub-Dep. of Astrophysics, Dep. of Physics, University of Oxford, Denys Wilkinson Building, Keble Road, Oxford OX1 3RH, United Kingdom\\
$^{2}$European Southern Observatory,  Karl-Schwarzschild-Stra\ss{}e 2, 85748, Garching, Germany\\
$^{3}$INAF -  Osservatorio Astronomico di Capodimonte, Via Moiariello  16, 80131, Naples, Italy\\
$^{4}$ Department of Physics, University of Warwick, Gibbet Hill Road, Coventry CV4 7AL, UK\\
$^{5}$ Université Côte d'Azur, Observatoire de la Côte d'Azur, CNRS, Laboratoire Lagrange, France \\
$^{6}$ V.N. Karazin Kharkiv National University, Sumska 35, Ukraine\\
$^{7}$ Institute of Radio Astronomy of National Academy of Science of Ukraine, Mystetstv 4, Ukraine\\
$^{8}$ Department of Astronomy and Space Informatics, V. N. Karazin Kharkiv National University, 35 Sumska Str., Kharkiv, Ukraine\\
$^{9}$ Instituto de Astrof\'isica de Canarias, V\'ia L\'actea s/n, E-38205 La Laguna, Tenerife, Spain\\
$^{10}$ Departamento de Astrof\'isica, Universidad de La Laguna, E-38200, La Laguna, Tenerife, Spain\\
$^{11}$ Finnish Centre for Astronomy with ESO, (FINCA), University of Turku, 20014 Turku, Finland \\   
$^{12}$ Tuorla Observatory, Department of Physics and Astronomy, University of Turku, 20014 Turku, Finland \\
$^{13}$ Turku Collegium for Science, Medicine and Technology (TCSMT), University of Turku, FI-20014 Turku, Finland \\
$^{14}$ School of Physics, H.H. Wills Physics Laboratory, Tyndall Avenue, University of Bristol, Bristol BS8 1TL, UK\\
}

\date{Accepted XXX. Received YYY; in original form ZZZ}

\pubyear{2026}

\begin{document}
\label{firstpage}
\pagerange{\pageref{firstpage}--\pageref{lastpage}}
\maketitle
\begin{abstract}
In this second paper of the \EINSPIRE\ series, we train a machine-learning–based regression on $\sim430$ nearby ($z<0.5$) ultra-compact massive galaxies (UCMGs) with spectroscopically inferred kinematics, stellar population parameters and a measured ``degree of relicness'' (DoR). Our goal is to investigate how robustly the spectroscopically inferred DoR can be statistically reconstructed from observable galaxy properties, and to explore the potential applicability of this framework to future wide-area surveys. We test several regression algorithms finding that Support Vector Regression (SVR) provides the best performance. We explore multiple input feature configurations, from a minimal set including only age and metallicity to more comprehensive ones incorporating stellar population parameters, kinematics, structural properties, and the associated uncertainties. All tested models achieve similarly high performance on the training set ($R^2\ge0.81$), except for the minimal configuration ($R^2\sim0.78$). When evaluated on an independent \INSPIRE\ sample of 52 UCMGs, the predictive power remains robust, although with increased model-to-model variation. The DoR distribution shows three regimes, with low (DoR$<0.3$) and high (DoR$>0.6$) values sparsely populated, leading to mild regression shrinkage toward intermediate values. However, this behaviour enables a conservative selection strategy: galaxies with predicted DoR$\ge0.6$ are strongly biased toward genuine extreme relics, making them prime targets for follow-up observations. This proof-of-concept confirms that the spectroscopically inferred DoR is robustly connected to observable stellar population and kinematical properties, and provides a first step toward future relic-candidate selection strategies in large photometric and spectroscopic surveys. 
\end{abstract}



\section{Introduction}
\label{sec:intro}
Relics \citep{Trujillo+09_superdense} are ultra-compact and massive galaxies (UCMGs, \citealt{Taylor+10_compacts, Poggianti+13, Tortora+16_compacts_KiDS, Tortora+18_UCMGs, Scognamiglio20}) 
that are almost exclusively made of “in-situ” very old stellar populations \citep{Ferre-Mateu+17, Spiniello20_Pilot, Spiniello+21, DAgo23, Spiniello24}.   
They are the local counterpart of high-$z$ "red nuggets" \citep{Damjanov+09}, the end product of the first phase of the so-called two-phase formation scenario \citep{Oser+10, Naab+14, Huertas-Company+16}. Given the stochastic nature of mergers, relics evolved passively and undisturbed after forming the bulk of their stars at very early cosmic times through a violent and quick starburst, and hence are local fossils of the early Universe.  
As such, they present an exciting opportunity to track the formation of the pristine stellar component, which is instead mixed with the accreted one in typical massive early-type galaxies (ETGs), affecting its spatial and orbital distributions.

Studying relics is also a very powerful way to indirectly put constraints on the size growth phase. In fact, simulations predict that the fraction of galaxies that do not grow in size is about 1 - 15\% \citep[e.g.,][]{Quilis_Trujillo13}, but this number highly depends on the physical processes acting during the second phase ($z<2$), and in particular on the relative contribution of major and minor mergers, gas accretion, and internal processes \citep[e.g.,][]{Bezanson09,Ownsworth14, Conselice22, Moura24}.

The INvestigating Stellar Populations In RElics survey has provided the first large catalogue of 38 spectroscopically confirmed relics in the near-by Universe \citep[][hereafter S24]{Spiniello24}.  With the Extension of the INvestigating Stellar Populations In RElics (\EINSPIRE) project, we now aim at enlarging the original \INSPIRE\  catalogue of UCMGs in redshift, stellar mass, and wavelength. In Paper I \citep[hereafter \EINSPIRE\ I]{Mills25}, we bridged the gap between the local Universe \citep{Ferre-Mateu+17, Yildirim17, Grebol2023} and the \INSPIRE\ catalogue of 52 UCMGs at $0.2<z<0.5$. We built a catalogue of 430 spectroscopically-confirmed UCMGs found in the Sloan Digital Sky Survey (SDSS) DR18 \citep{Almeida+23_SDSS} at redshifts $z<0.3$. To separate younger UCMGs that have yet to start the second phase of mass assembly from relics, the \textit{degree of relicness} (DoR, \citealt{Ferre-Mateu+17}), was defined in \citet{Spiniello+21}, and subsequently used in all \INSPIRE\ and \EINSPIRE\ papers. The DoR is defined as a dimensionless number ranging from 0 to 1, quantifying how "extreme" the SFH of an object is\footnote{Specifically, the DoR is operationally defined as the re-normalised mean of three quantities: the stellar mass fraction formed by $z=2$, the inverse of the cosmic time at which 75\% of the mass is assembled, and the inverse redshift-normalised final assembly time (see Eq.~1 in \citetalias{Spiniello24}).}. A high DoR indicates an early complete mass assembly, with almost no contribution from later star formation episodes. A lower DoR instead indicates that there is a non-negligible percentage of populations with younger ages, and hence a much later time of final assembly. 
Currently, the highest DoR measured from observations is held by the local relic NGC~1277 \citep{Trujillo+14, Ferre-Mateu+17} with DoR$=0.94$. A UCMG characterised by a time-extended SFH that is still forming a small percentage of stars would instead have DoR $\sim0$. Interestingly, in \citetalias{Mills25} we found a strong correlation between the DoR and the stellar velocity dispersion, metallicity and [Mg/Fe] (in agreement with previous results, e.g. \citealt{DAgo23,Grebol2023, Spiniello24}). 
The importance of high-spatial-resolution imaging has recently been demonstrated for KiDS~J0842+0059, the first fully confirmed relic beyond the local Universe: adaptive-optics imaging verified its $\sim1$ kpc compact structure and showed that its stellar-mass density profile closely resembles those of NGC~1277 and high-redshift red nuggets \citep{Tortora25}.

Given their small sizes and low number densities, \footnote{The relics number density has been inferred to be $6\times10^{-7}$ Mpc$^{-3}$ at $z\sim0$ \citep{Ferre-Mateu+17} and ranging between $10^{-4}$ and $10^{-6}$ Mpc$^{-3}$ up to $z\sim0.5$ \citep{Spiniello+21}.} detecting and studying a statistically valid sample of relics at different cosmic epochs requires a very large sky coverage and images with exquisite spatial resolution and good sensitivity. This is exactly what the ESA’s Euclid mission is starting to provide \citep{Mellier24}. In roughly 6 years of operations, the Euclid’s Wide Survey \citep{Scaramella22}, will cover $\sim14000$ deg$^2$, corresponding to more than a third of the total sky, observing billions of galaxies out to redshift $z\sim2$ with an unprecedented spatial resolution, precision and sensitivity. 
Euclid will allow us to find and count UCMGs across the entire size-growth epoch, collecting large and statistically significant high-purity samples. 
The mission will also provide near-infrared (NIR) spectroscopy allowing the inference on the stellar velocity dispersion, but unfortunately, it has a too poor spectral resolution to enable a detailed stellar population analysis. An alternative path is obtaining spectral energy distribution (SED) from the multi-band photometric images, computing integrated and spatially-resolved SFRs, ages and metal abundances \citep{Enia25,Nersesian25,Kovacic25,Abdurrouf25}. Although it does not provide an ultimate confirmation of the relic's nature, this information can be used to select the most reliable and interesting objects for high-resolution spectroscopic follow-ups.

In this second paper of the \EINSPIRE\ series, we investigate how robustly the spectroscopically inferred DoR can be statistically reconstructed from commonly derived stellar population, kinematical, and structural observables across independent UCMG samples. The goal of the present work is therefore not to replace the physically motivated spectroscopic determination of the DoR, nor to claim that machine learning uncovers fundamentally new physical quantities inaccessible to traditional analyses. Rather, we aim to test the stability, predictability, and generalisability of the empirical relations linking the DoR to observable galaxy properties. To this end, we implement and compare several machine-learning regression frameworks trained on spectroscopically characterised UCMGs from the \EINSPIRE\ survey and validated on the independent \INSPIRE\ sample. Since the input quantities themselves are derived from stellar population analyses, the present implementation should be interpreted primarily as a proof-of-concept investigation of how the relic signal manifests across observable parameter space, rather than as a fully independent estimator of relicness. Nevertheless, the analysis provides a quantitative way to identify which observables retain the strongest predictive connection with the DoR, how stable these relations remain under moderate distributional shifts between datasets, and which regions of parameter space naturally correspond to the most extreme relic systems.

The broader motivation for this approach is that future large-area surveys such as Euclid, the Dark Energy Spectroscopic Instrument (DESI) survey \citep{DESI_I_2025}, and the 4MOST survey \citep{4MOST19}, will dramatically increase the number of known UCMGs while simultaneously providing heterogeneous observational information of varying quality and completeness. Although the present implementation still relies on spectroscopically derived stellar population quantities, understanding which measurable galaxy properties most strongly encode the relic signal is an important first step toward developing scalable strategies for prioritising high-probability relic candidates for detailed spectroscopic follow-up.

The paper is organised as follows. We begin in Section~\ref{sec:data} by giving a short overview of the data used in this paper. In Section~\ref{sec:ML} we describe the machine learning (ML) method we use to predict the DoR of the UCMGs, along with the tests we run on the input data. The ML results are presented and discussed in Section~\ref{sec:results}, displaying how the data-driven approach performs on both the training and the test datasets.   
We finally discuss our results and their implications for future UCMGs searches in Section~\ref{sec:discussion} and summarise the findings in Section~\ref{sec:conclusions}.

Throughout the paper, we assume a standard $\Lambda$CDM cosmology with $H_0=67.7$ \kms Mpc$^{-1}$, $\Omega_{\mathrm{\Lambda}}=0.689$, and $\Omega_{\mathrm{M}}=0.311$ \citep{Planck+20}.

\section{Data}
\label{sec:data}

\begin{figure*}
    \centering
    \includegraphics[width=0.99\linewidth]{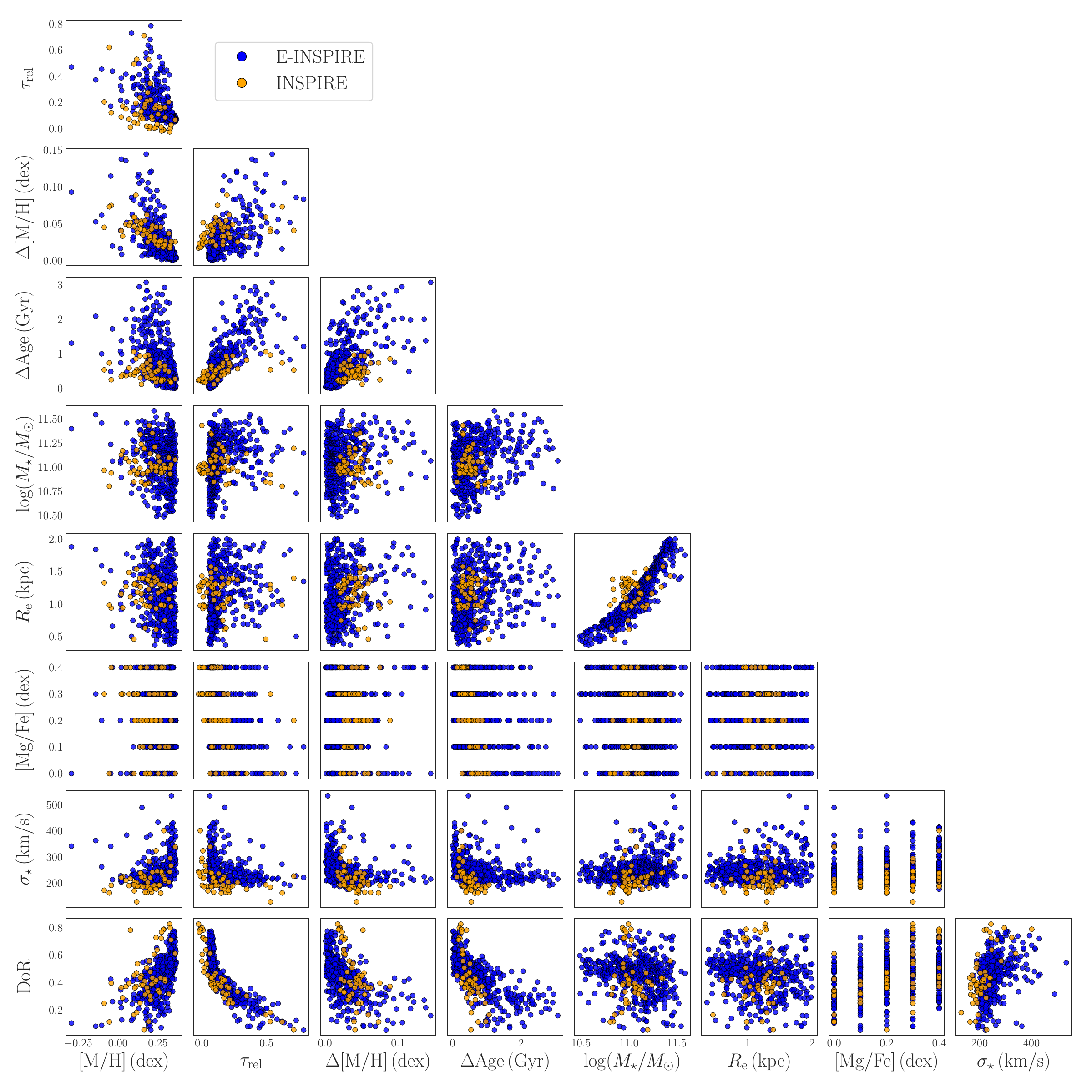}
    \caption{Corner Plot of the stellar populations, kinematics and sizes parameters, alongside their uncertainties ($\Delta$) and the computed DoR, for the \INSPIRE\ (orange) and \EINSPIRE\ (blue) UCMGs.}
    \label{fig:corner_plain}
\end{figure*}

\begin{figure*}
    \centering
    \includegraphics[width=0.99\linewidth]{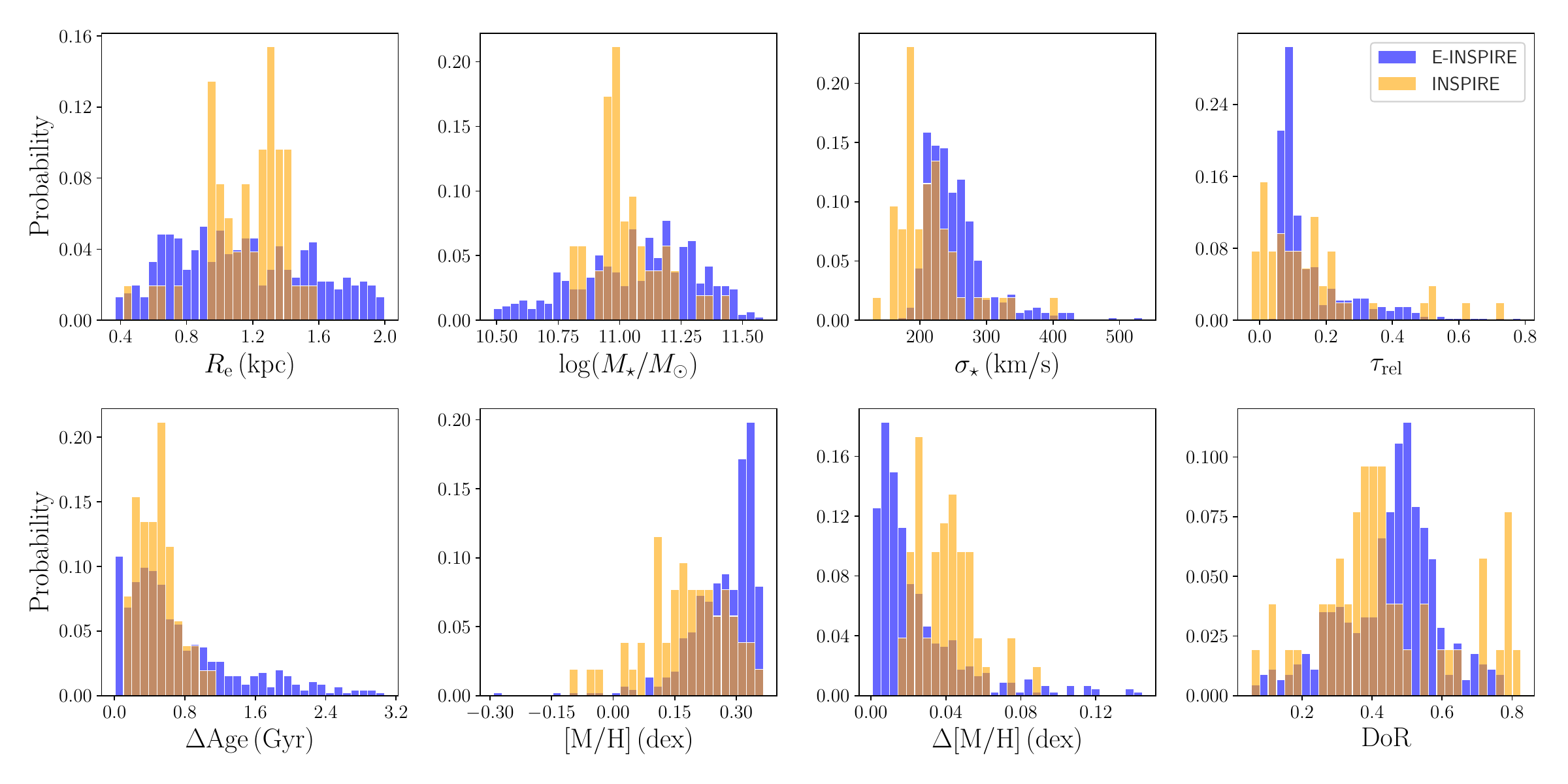}
    \caption{
        Histograms showing the \EINSPIRE\ (blue)  training data and the \INSPIRE\ (orange)  testing data, scaled to allow direct comparison.
    }
    \label{fig:histogram}
\end{figure*}

The data used in this paper comes from the \INSPIRE\ and \EINSPIRE\ surveys. In particular, we use the \EINSPIRE\ larger catalogue of UCMGs presented in \citet{Mills25} as a training set for the ML algorithm. This dataset comprises a total of 430 galaxies in the redshift range $0.03<z<0.3$ with stellar masses larger than $3\times10^{10}$ \Msun, stellar velocity dispersion $\sigma_{\star}>130$ \kms and compactness $\log\Sigma_{1.5}>10.5$ \citep{Baldry21}.  We then test the ML algorithm on the smaller \INSPIRE\ set of 52 UCMGs selected from the KiDS Survey \citep{Kuijken19_KIDSDR4} in \citet{Tortora+16_compacts_KiDS, Tortora+18_UCMGs} and \citet{Scognamiglio20}, which was presented in \citetalias{Spiniello24}. This latter dataset has a slightly more conservative cut in stellar mass ($6\times10^{10}$ \Msun),  a fix cut in (circularised) effective radii (\Reff$<2$ kpc)\footnote{The mass and size threshold used to define UCMGs vary from paper to paper, sometimes in an independent way, sometimes combined into a pseudo-density measurement. We refer the readers to \citet{Charbonnier+17_compact_galaxies} for a review of the various compactness criteria used in the literature.} and covers slightly higher redshifts: $0.1<z<0.5$. For each of the UCMGs (in both \INSPIRE\ and \EINSPIRE), full-spectral fitting via single stellar populations (SSPs) modelling was performed in previous papers. Hence stellar ages, metallicities ([M/H]) and [Mg/Fe] ratios as well as star formation histories (SFHs), metallicity evolution histories (MEHs) and DoR estimates are available. We refer the reader to \citetalias{Spiniello24} and \citetalias{Mills25} for a detailed description of the two surveys. 

Here we focus on the main differences between the two datasets. First, the UCMGs are identified with two different criteria in the two surveys. While the \EINSPIRE\ objects have been selected using the compactness criterion by \citet{Baldry21}, the \INSPIRE\ ones have been selected with fixed radius and stellar mass thresholds of R$_{\rm e} \le 2$ kpc and M$_{\star} > 6 \times 10^{10}$ M$_{\odot}$. 

Another substantial difference between the two datasets is the metallicity range of the MILES SSP models \citep{Vazdekis15} used in the full-spectral fitting procedure. Indeed, \citetalias{Mills25} adopt a metallicity range limited to [M/H]$\le0.26$ dex, while \citetalias{Spiniello24} extend the grid to higher metallicities, up to [M/H]$=0.40$ dex. These extensions are provided by the MILES team via extrapolation beyond the original model parameter space and are explicitly labelled ``unsafe'' by the model developers. The extended range was adopted in \INSPIRE\ because relic galaxies, and particularly the most extreme ones, were systematically reaching the upper metallicity boundary when the fit was restricted to [M/H]$\le 0.26$. As shown in Fig.~A6 of \citetalias{Mills25}, including the higher [M/H] models leads to systematically lower DoR estimates, but leaves the overall shape of the DoR distribution broadly unchanged. Here, we remove the upper metallicity restriction in order to construct a homogeneous set and to avoid artificial boundary effects in the spectroscopic fits, particularly for the most extreme relics. This choice also enables a more consistent comparison with future photometric SED-based analyses, where metallicity constraints are typically weaker. Hence, before proceeding with our ML algorithm, we re-derive all the SFHs for the \EINSPIRE\ I objects by running the code Penalised Pixel-fitting\footnote{\url{https://pypi.org/project/ppxf/}} (\ppxf; \citealt{Cappellari04,Cappellari17,Cappellari23}) using the un-restricted set of models. Allowing the fit to use the highest metallicity models, an expected shift to higher metallicity and lower age is found for some galaxies, due to the well-known metallicity-age degeneracy (\citealt{Worthey+94, Worthey+99}). Even extending the models to [M/H]$=0.40$ dex, we found that the majority of the \EINSPIRE\ objects with DoR$>0.6$ hit the highest metallicity limit, hinting for a direct and strong correlation between DoR and [M/H], stronger than the suggested metallicity-compactness relation \citep{Beverage21}. 

Another important aspect to consider when comparing the \INSPIRE\ and \EINSPIRE\ datasets is that they cover a different redshift range. Clearly, higher redshift galaxies cannot be as old in absolute terms as local galaxies (simply because of the younger age of the Universe at which they are observed). Hence, we define instead a redshift-normalised age quantity: 
\begin{equation}
\tau_{\rm rel} = \frac{t_{\rm Uni}-t_{\rm gal}}{t_{\rm Uni}}
\end{equation}
where $t_{\rm gal}$ is the age of each object and $t_{\rm Uni}$ is the age of the Universe at the redshift of any given object. 

Finally, we note that while age and metallicity are derived via full spectral fitting, [Mg/Fe] is estimated from index--index diagrams (see \citetalias{Spiniello24} and \citetalias{Mills25} for details), using a model grid with a nominal step of 0.1 dex. Although the model grids can in principle be interpolated to a finer resolution, the typical uncertainties on the measured line indices limit the effective precision of the [Mg/Fe] determination to a comparable level. As a result, it is more difficult to draw firm conclusions regarding the presence of subtle correlations between [Mg/Fe] and DoR. 

Figure~\ref{fig:corner_plain} shows the distribution and correlations between the above described parameters, and their statistical uncertainties (denoted with $\Delta$) when available, for both the \INSPIRE\ (orange) and \EINSPIRE\ I (blue) galaxies. Several parameters correlate with DoR. However, this correlation changes for different DoR ranges. For instance, the [M/H] is positively correlated with DoR until around DoR$\sim0.6$, but then becomes less efficient in distinguishing massive relic galaxies above this DoR threshold. In a similar way, $\tau_{\rm rel}$ and DoR are tightly inversely correlated, as expected, but only for $\tau_{\rm rel}$ of $> 0.2$. Below this value,  $\tau_{\rm rel}$ is not a useful discriminator for DoR values (i.e. above DoR$\sim0.6$). Another interesting feature is that for very high metallicity and very small $\tau_{\rm rel}$ values, the uncertainties on age and [M/H] are generally always small, while they increase dramatically for lower DoRs ($\sim0.4$). Previous results have shown that relics often have a higher velocity dispersion \citep{DAgo23, Grebol2023, Spiniello24, Mills25}. The plot shows that at higher DoR, compact galaxies span a wide range of $\sigma_{\star}$\footnote{Extracted from an aperture encapsulating 50\% of the light from \INSPIRE\ and from the entire SDSS spectrum from \EINSPIRE.}, while there are very few points with DoR$<0.3$ and $\sigma_{\star}>300$ km/s. Finally, for both datasets, stellar masses and effective radii do not show any correlation with DoR, however these measurements are affected by large uncertainties since calculated from ground-based observations \citep{Tortora+16_compacts_KiDS}.

In machine-learning applications, predictive performance may deteriorate when the test data occupy regions of feature space that are poorly represented by the training data. We therefore compared the INSPIRE and E-INSPIRE samples using both statistical tests and direct inspection of their feature distributions. Two-sample Kolmogorov--Smirnov tests indicate statistically significant differences for all structural, kinematical, and stellar-population parameters ($D\sim0.2$--$0.5$, with $p\ll0.01$), \footnote{except $[\mathrm{Mg/Fe}]$} reflecting the different selection criteria, redshift coverage, and stellar-population modelling adopted by the two surveys. The two datasets should therefore not be regarded as random realisations of a single parent distribution. However, statistical differences do not imply complete separation of the feature space. As illustrated in Figures~\ref{fig:corner_plain} and~\ref{fig:histogram}, the two surveys still occupy regions of feature space with substantial quantitative overlap, as quantified below. 

We quantify this overlap by measuring, for each feature, the fraction of \INSPIRE\ galaxies falling within the 5th--95th percentile range of the corresponding \EINSPIRE\ distribution, together with the normalised mean offset $\delta \equiv (\mu_{\rm test} - \mu_{\rm train})/\sigma_{\rm train}$. Five of the nine quantities achieve overlap fractions of 98--100 per cent. Among these, $[\mathrm{Mg/Fe}]$ is the only feature showing no statistically significant distributional difference ($D = 0.09$, $p = 0.82$). The two features with the lowest overlap are $\sigma_\star$ (48 per cent, $\delta = -0.83$) and $\tau_{\rm rel}$ (60 per cent, $\delta = -0.06$). 
For $\sigma_\star$, the offset may reflect the instrumental difference between surveys; this feature is found to affect results minimally. For $\tau_{\rm rel}$, the most predictive feature, the 60 per cent overlap reflects a difference in distribution shape arising from the cosmological compression of $\tau_{\rm rel}$ at high redshift, while $\delta = -0.06$ confirms that the two surveys occupy essentially the same central values for this feature.

In conclusion, although the two datasets are not statistically identical, we intentionally use \INSPIRE\ as an external validation sample to assess whether the empirical relation between observable galaxy properties and DoR transfers across independently constructed UCMG samples, rather than only to randomly held-out galaxies drawn from the same distribution.

\section{METHODS: MACHINE LEARNING REGRESSION}
\label{sec:ML}
The increasing volume of data from wide-sky spectroscopic and photometric surveys (such as SDSS, \citealt{Almeida+23_SDSS}, the Dark Energy Survey, DES, \citealt{Abbott18, Abbott21}; Euclid \citealt{Mellier24}; and soon LSST, \citealt{LSST19}; and 4MOST, \citealt{4MOST19}),  enables the application of ML  techniques to problems of galaxy evolution \citep[e.g.][]{Ball10, VanderPlas12, Baron19, Martin20, Sharma20, Khramtsov21, Ofman22, Shao22, Humphrey22, Huertas23, Ciprijanovic23, Jones24, Enia24, Zhao25, Walmsley25, Siudek25, Kovacic25}. 

Because the DoR is defined as a continuous parameter bounded between 0 and 1, we formulate its estimation as a supervised regression problem rather than a binary classification task. The models are trained using galaxy observables as predictors and the spectroscopically inferred DoR as the ground-truth target. 
The training set is built from the 430 galaxies belonging to the \EINSPIRE\ survey, while the independent test set comprises the 52 galaxies from \INSPIRE. 

The choice of \EINSPIRE\ as the training set is primarily motivated by its substantially larger sample size, providing broader coverage of the observable parameter space and improving the stability of the regression models during optimisation. The \INSPIRE\ survey is retained as an independent external validation sample. As discussed in Section~2, the two surveys differ in selection criteria, redshift coverage, and stellar-population modelling, such that the validation evaluates the transferability of the learned mapping under moderate covariate shift rather than interpolation within a single statistical population. 
This represents a more demanding test than a random train--test split, as it preserves the observational differences between independently constructed surveys rather than distributing them across both subsets. Consequently, successful performance on the \INSPIRE\ sample suggests that the learned mapping transfers across independently selected UCMG catalogues rather than merely interpolating within a single statistical population. Such a setting also more closely reflects the intended application of the framework, where models trained on one spectroscopic sample will ultimately be applied to future galaxy surveys characterised by different selection functions and observational strategies. Despite these differences, the degradation in predictive performance between training and testing remains modest (Section~\ref{sec:results}), indicating that the empirical relation linking observable galaxy properties and DoR remains stable across independently selected UCMG samples.

The workflow consists of three steps: defining the feature sets, comparing regression algorithms, and evaluating predictive performance using cross-validation and an independent test dataset.


\subsection{Input features}
\label{sec:input_features}
The predictive framework relies on a set of observable galaxy properties derived from full spectral fitting and structural measurements. 
Our goal is two-fold: (i) to identify the most informative combination of features for recovering DoR, and (ii) to ensure flexibility when applying the method to datasets providing different subsets of observables.

Candidate input features span three physical categories: stellar populations, kinematics, and structural parameters. The stellar population quantities include the redshift-normalised age $\tau_{\rm rel}$, metallicity [M/H], and $\alpha$-abundance [Mg/Fe]. The kinematic information is represented by the stellar velocity dispersion $\sigma_{\star}$. Structural parameters include the stellar mass $\log M_{\star}$ and effective radius $R_{\rm e}$.

In addition, we test the inclusion of the uncertainties on the stellar population parameters ($\Delta \mathrm{age}$ and $\Delta$[M/H]) as input features. These quantities may carry physical information, particularly in the case of extreme relics. 
If such systems are dominated by relatively homogeneous old stellar populations, the spectral fitting solution may become more stable against variations in template combinations, potentially leading to smaller formal uncertainties in age and metallicity. However, these uncertainties are not purely physical quantities. They also depend on observational factors such as signal-to-noise ratio, spectral coverage, and modelling degeneracies inherent to the SSP fitting procedure. In this work, we therefore treat $\Delta {\rm Age}$ and $\Delta [{\rm M/H}]$ primarily as empirical descriptors of the stability of the recovered stellar population solution rather than as fully independent physical observables. 
In this scenario, the stability of the solution reflects the homogeneity of the underlying stellar population. This argument does not apply to the uncertainty on the velocity dispersion, which is instead primarily driven by the signal-to-noise ratio of the input spectrum rather than by population complexity (see \citealt{DAgo23}).

While stellar population and kinematic parameters have been shown to correlate with relic properties, effective radius and stellar mass are not expected to provide strong discriminating power within a sample already restricted to UCMGs. Nevertheless, we test their inclusion for completeness and for future applicability to datasets where structural parameters may be measured with higher precision (e.g. from space-based imaging).

For models whose behaviour depends on absolute feature magnitudes,  
maintaining consistent feature scaling between the training and deployment datasets is essential. Indeed, some of the algorithms we will use rely on distance-based calculations in the feature space via the kernel function (the chosen distance measure). This means that features with larger magnitudes would dominate the optimisation if left unscaled. We therefore apply standardisation (zero mean, unit variance) to all input features, fitting the scaler on the training data and applying the same transformation to the test data to avoid information leakage and to preserve consistency in the feature space. 

Table~\ref{tab:SVR_MODELS} summarises all tested feature configurations. We begin with the minimal set composed of $\tau_{\rm rel}$ and [M/H] only (Model ID 1). Model ID 2 extends this baseline by including their associated uncertainties. Model ID 3 further incorporates the stellar velocity dispersion $\sigma_{\star}$, motivated by the observed correlation between relicness and kinematics (\citealt{DAgo23}; \citetalias{Spiniello24}). Model ID 4 instead includes the $\alpha$-abundance [Mg/Fe], while Model ID 5 combines both $\sigma_{\star}$ and [Mg/Fe].

However, models including $\sigma_{\star}$ and [Mg/Fe] (IDs 3–5) may not be directly applicable to the Euclid Wide survey \citep{Scaramella22}, as these quantities require high-resolution spectroscopy over wavelength ranges that are not accessible with Euclid’s slitless near-infrared spectroscopic data. 
We therefore define a configuration based on quantities that are expected to become approximately accessible through future survey-level photometric or SED-fitting analyses, namely stellar population parameters (age and metallicity), their uncertainties, and structural quantities ($\log M_{\star}$ and $R_{\rm e}$), labelled as the `Euclid-like' model (ID 6).

Model ID 7 extends this configuration by reintroducing $\sigma_{\star}$, in the event that Euclid spectra prove sufficient for secure velocity dispersion measurements. Finally, Model ID 8 (`Complete Set') includes all available features simultaneously. 

Depending on the configuration, the dimensionality of the feature space ranges from 2 to 8.
\subsection{Algorithm comparison and selection}

\begin{figure}
    \centering
    \includegraphics[width=0.99\linewidth]{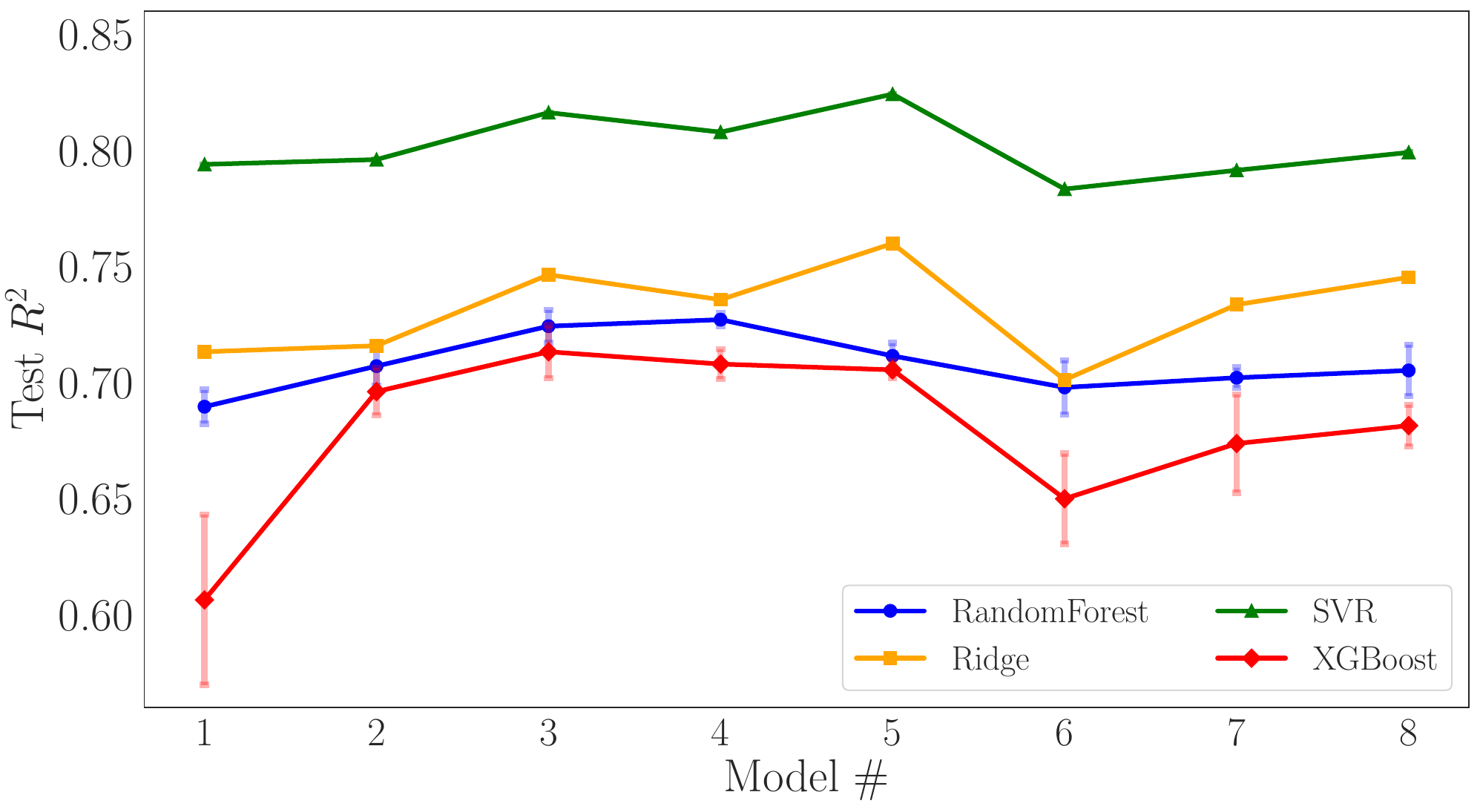}
    \caption{The test performance of the different algorithms for each of the model feature sets with error bars showing the standard deviation across different random seeds. SVR and Ridge Regression are deterministic and therefore show no variation across seeds.}
    \label{fig:algos}
\end{figure}

To assess the robustness of the mapping between galaxy observables and DoR, we tested four conceptually distinct regression families: linear, tree-based ensemble, and kernel-based methods. This comparison allows us to determine whether the relation is primarily linear or intrinsically non-linear within the explored feature space. All algorithms were implemented using the \texttt{scikit-learn} library \citep{Pedregosa11}.

The linear baseline is tested through Ridge Regression (RR; \citealt{Hoerl1970}). RR provides a linear model with $\ell_2$ regularisation, assuming a linear relationship between the input features and DoR while penalising large coefficients to mitigate multicollinearity and overfitting. Given the known correlations among stellar population parameters (e.g. the age--metallicity degeneracy; \citealt{Worthey+99}), Ridge is particularly suitable as a stable and interpretable reference model. However, its linear functional form limits its ability to capture higher-order interactions between parameters.

Tree-based methods are attractive because they model complex non-linear decision boundaries without requiring explicit feature transformations. However, they may be less stable when extrapolating beyond the training distribution and can exhibit higher variance across cross-validation folds for relatively small datasets. As tree-based ensemble approaches, we tested both Random Forest (RF; \citealt{breiman2001random}) and XGBoost \citep{Chen16}. Random Forest constructs an ensemble of decision trees trained on bootstrapped samples and naturally captures non-linear interactions while remaining robust to noise and moderate overfitting. XGBoost is a gradient-boosted tree ensemble that sequentially improves residual errors; it is computationally efficient, handles missing values natively, and often achieves state-of-the-art performance on structured tabular data.

Finally, we test Support Vector Regression (SVR; \citealt{drucker1997}), a kernel-based method that performs linear regression in a transformed high-dimensional feature space defined implicitly by a kernel function. Using a polynomial kernel, SVR captures smooth non-linear feature interactions while retaining explicit regularisation/kernel control through the hyperparameters $C$, $\epsilon$, and $\gamma$. 
SVR is particularly well suited to problems characterised by moderate sample sizes, relatively high-dimensional feature spaces, and smooth, structured non-linear mappings. Compared to tree-based methods, SVR enforces a more global and smoother functional form, which is advantageous when the underlying physical relation is expected to be continuous rather than piecewise.

Model performance is quantified using the coefficient of determination, $R^2$:

\begin{equation}
R^2 = 1 - \frac{\sum_{i=1}^{n} (y_i - \hat{y}_i)^2}
{\sum_{i=1}^{n} (y_i - \bar{y})^2},
\end{equation}

where $y_i$ and $\hat{y}_i$ denote the true and predicted DoR values, respectively, and $\bar{y}$ is the mean of the observed values. We additionally report the root-mean-square error (RMSE) when discussing prediction quality. 
To obtain a robust estimate of the training performance, we adopt $k$-fold cross-validation ($k=5$) on the \EINSPIRE\ sample. The average $R^2$ across folds is reported as $R^2_{\rm training}$, with its standard deviation reflecting the sensitivity to the data split. Performance on the \INSPIRE\ sample is reported as $R^2_{\rm test}$.

Figure~\ref{fig:algos} compares the performance when trained on \EINSPIRE\ and tested on \INSPIRE\, quantified by the coefficient of determination ($R^2_{\rm test}$), of the four algorithms across the different feature configurations described in Section~\ref{sec:input_features}. SVR consistently achieves the highest predictive performance. Given this stability, we adopt SVR for the remainder of the analysis. It is also well known to be less prone to overfitting relative to tree-based models. 

SVR predicts the DoR for a new galaxy $\mathbf{x}$ as

\begin{equation}
\hat{y}(\mathbf{x}) = \sum_{i=1}^{N} (\alpha_i - \alpha_i^*) \, K(\mathbf{x}_i, \mathbf{x}) + b,
\end{equation}

\noindent where $\mathbf{x}_i$ are the training feature vectors, $N$ is the number of training samples, $K(\mathbf{x}_i,\mathbf{x})$ is the kernel function, $\alpha_i$ and $\alpha_i^*$ are the Lagrange multipliers obtained from the dual optimisation problem, and $b$ is the bias term. Only the support vectors (i.e. those training points for which $\alpha_i - \alpha_i^* \neq 0$) contribute to the sum.

We employ a polynomial kernel,

\begin{equation}
K(\mathbf{x}_i, \mathbf{x}_j) = (\gamma \, \mathbf{x}_i \cdot \mathbf{x}_j + r)^d,
\end{equation}

\noindent where $\gamma$ controls the scaling of the feature space, $d$ is the polynomial degree, and $r$ is a constant offset added before exponentiation. A non-zero $r$ allows lower-order interactions to remain relevant even when higher polynomial degrees are used, thereby regulating the balance between linear and higher-order contributions. 
We adopt here a polynomial degree $d=3$, which provides sufficient flexibility to capture non-linear feature interactions while maintaining stability across cross-validation folds.

\begin{table}
\centering
\caption{Adopted SVR hyperparameters.}
\begin{tabular}{lcl}
\hline
Parameter & Value & Role \\
\hline
$C$ & 5.0 & regularisation strength (bias--variance trade-off) \\
$\epsilon$ & 0.03 & insensitive loss width (noise tolerance) \\
kernel & poly & polynomial kernel \\
$\gamma$ & 0.01 & feature-space scaling \\
degree & 3 & polynomial order \\
$r$ & 0.5 & kernel offset term \\
\hline
\end{tabular}
\label{tab:model-params}
\end{table}

\begin{table*}
\centering
\caption{
The performance of the different set of SVR models with different features, on both training and test data.}
\label{tab:SVR_MODELS}
\begin{tabular}{lccc}
\hline
Model ID & Input Features & R$^2_{\rm training}$ & R$^2_{\rm test}$ \\
\hline
1 - Stel. pop. & [M/H], $\tau_{\rm rel}$ & 0.777 $\pm$ 0.005 & 0.794  \\
2 - Stel. pop. with errors & [M/H], $\tau_{\rm rel}$, $\Delta$[M/H], $\Delta$age & 0.820 $\pm$ 0.004 & 0.796  \\
3 - Stel. pop. and kinematics & [M/H], $\tau_{\rm rel}$, $\Delta$[M/H], $\Delta$age, $\sigma_\star$ & 0.826 $\pm$ 0.003 & 0.816  \\
4 - Stel. pop. and $\alpha$-abundance & [M/H], $\tau_{\rm rel}$, $\Delta$[M/H], $\Delta$age, [Mg/Fe] & 0.823 $\pm$ 0.004 & 0.808 \\
5 - Stel. pop., $\alpha$-abundance and kinematics & [M/H], $\tau_{\rm rel}$, $\Delta$[M/H], $\Delta$age, [Mg/Fe], $\sigma_\star$ & 0.828 $\pm$ 0.002 & 0.824  \\
6 - Stel. pop. and structural (Euclid-like) & [M/H], $\tau_{\rm rel}$, $\Delta$[M/H], $\Delta$age, $\log$ M$_\star$, R$_{\rm e}$ & 0.818 $\pm$ 0.003 & 0.783  \\
7 - Stel. pop., structural and kinematics & [M/H], $\tau_{\rm rel}$, $\Delta$[M/H], $\Delta$age, $\log$ M$_\star$, R$_{\rm e}$, $\sigma_\star$ & 0.823 $\pm$ 0.003 & 0.791  \\
8 - Complete Set & [M/H], $\tau_{\rm rel}$, $\Delta$[M/H], $\Delta$age, $\log$ M$_\star$, R$_{\rm e}$, [Mg/Fe], $\sigma_\star$ & 0.827 $\pm$ 0.003 & 0.799  \\
\hline
\end{tabular}
\end{table*}

Hyperparameters were optimised through a grid search over logarithmically spaced values of $C$, $\gamma$, and $\epsilon$, with performance evaluated using five-fold cross-validation on the training set. The adopted hyperparameters are summarised and explained in Table~\ref{tab:model-params}. 
Variations around these values lead to only marginal changes in performance ($\Delta R^2 \lesssim 0.01$, even for order-of-magnitude variations), indicating that the model operates within a stable region of parameter space.



\section{RESULTS}
\label{sec:results}

The performance on the training (testing) sample is shown in the third (fourth) column of Table~\ref{tab:SVR_MODELS}. 
As expected from previously discussed factors, a reasonable deviation is found between the performance in the two datasets. To verify whether this gap is due to overfitting, we tested varying the SVR hyperparameters to reduce model complexity (e.g., decreasing the regularisation parameter $C$, increasing $\epsilon$, and reducing the polynomial degree). These restrictions caused both training and testing performance to suffer similarly. This suggests that the performance difference stems from distributional differences between the datasets rather than model overfitting. 


Surprisingly, there is little variation in the model performance across feature configurations. This likely reflects both intrinsic correlations among the input parameters and the restricted diversity of the training sample. While the tested features encode physically distinct quantities, many of them are correlated for relic systems. In particular, the most extreme massive relics identified in \citetalias{Spiniello24} occupy a very narrow region of parameter space, characterised by high velocity dispersion, high metallicity and enhanced [Mg/Fe] with minimal scatter. Such intrinsic clustering reduces the marginal gain obtained by adding further correlated features, as the relevant information is already captured by a limited subset of parameters. Given the relatively low standard deviation in prediction across cross-validation folds, the model is also fairly insensitive to the specific subset of data on which it is trained.


\begin{figure}
    \centering    
    \includegraphics[width=0.99\linewidth]{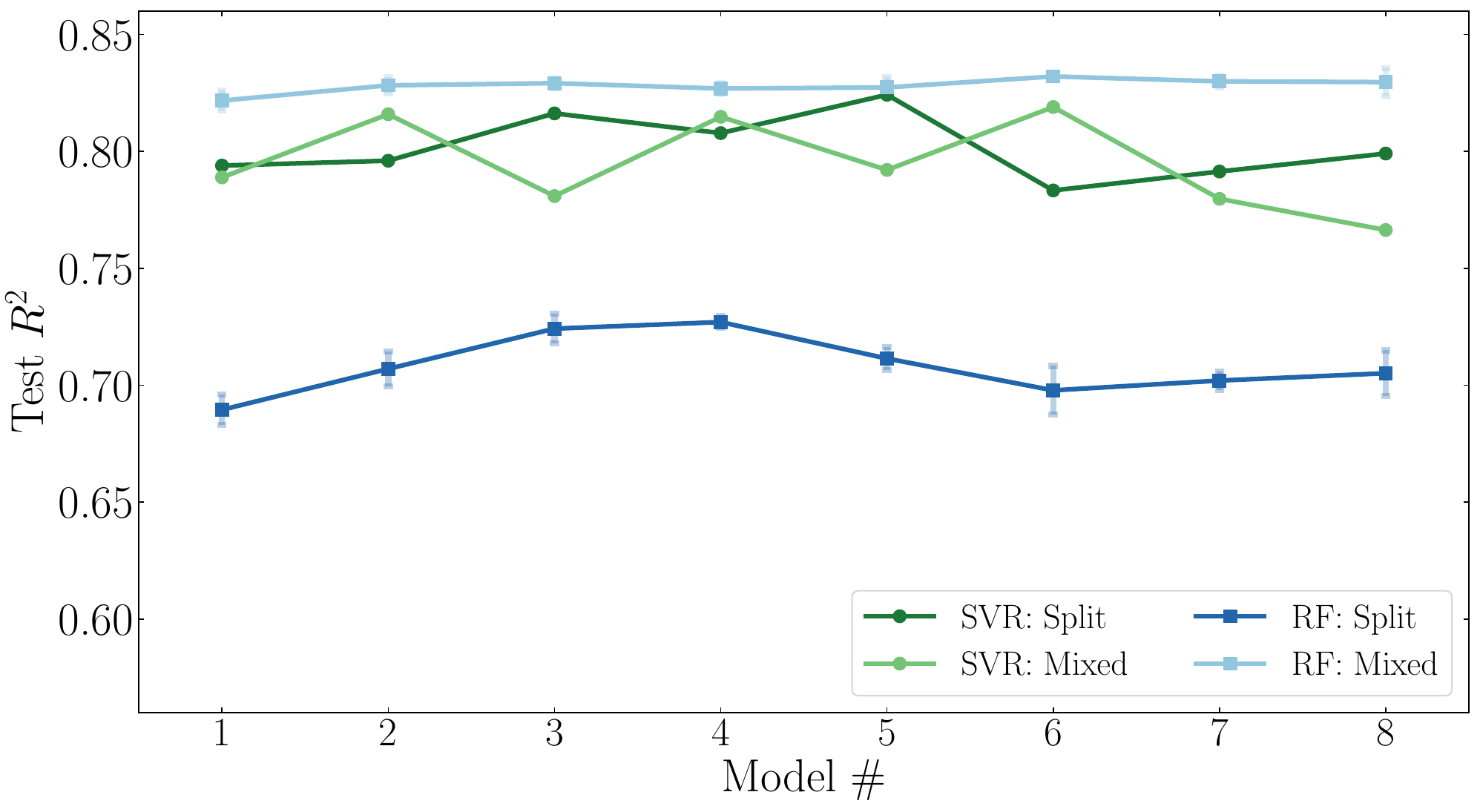}
    \caption{SVR and RF model performance per model for both mixed (SVR - light green; RF - light blue) data and models trained on \EINSPIRE\ and tested on \INSPIRE\ data (SVR - dark green; RF - dark blue). }
    \label{fig:mixed_domains_comparison}
\end{figure}

On the independent \INSPIRE\ test sample, performance is slightly lower than for the training sample but remains high. The lowest-performing configuration (Model~6) yields $R^2 = 0.783$, whereas Model~5, which combines stellar population parameters, $\alpha$-abundance, and kinematics, achieves the highest performance with $R^2 = 0.824$. This highlights the value of incorporating as much spectroscopic information as possible, where available, when prioritising high-confidence relic candidates. Once again, we stress that a model explaining $\sim78$\% of the variance in an entirely independent test sample is sufficient for assembling a large catalogue of reliable relic candidates. More importantly, the regression model retains strong predictive performance when applied to the independently selected \INSPIRE\ sample, supporting the transferability of the learned mapping beyond the specific selection function of the \EINSPIRE\ training sample.

As an additional robustness test, we repeated the analysis after pooling the \INSPIRE\ and \EINSPIRE\ samples and randomly redefining the training and testing subsets (Figure~\ref{fig:mixed_domains_comparison}). The resulting SVR performance remained nearly unchanged with respect to the original external-validation experiment, while Random Forest showed a significant improvement. The main conclusions therefore remain insensitive to the adopted train--test definition, further supporting the stability of the learned relation between observable galaxy properties and DoR recovered by SVR. Overall, the performance varies only modestly across feature configurations, with $R^2$ values spanning a narrow range between $0.783$ and $0.824$. Importantly, the lower performance of the structural-parameter configurations (Models~6--8) persists under both validation strategies, indicating that this behaviour is intrinsic to the predictive content of those features rather than a consequence of the original train--test definition.

In the following, we illustrate the behaviour of the regression framework by comparing the best- and worst-performing models on the test sample, thereby bracketing the range of achievable performance under different feature assumptions. The lowest-performing configuration corresponds to the Euclid-like setup, designed to mimic a reduced-feature survey scenario. This configuration should be interpreted as a proof-of-concept for relic selection under restricted inputs rather than as a direct assessment of Euclid survey products. In particular, it relies on spectroscopically derived ages and metallicities as proxies for SED-based quantities, which differ in uncertainty and systematics from true survey measurements. We retain this configuration as a controlled test of how the regression behaves under restricted features matching the coverage (though not the precision) of future SED-based Euclid measurements. While the Euclid-like results demonstrate encouraging robustness of the method under simplified inputs, definitive conclusions for Euclid itself will require retraining and validation on survey-calibrated parameters and end-to-end simulations. 

\subsection{Predicted DoR and turning points}
\begin{figure*}
    \centering
    \includegraphics[width=\linewidth]{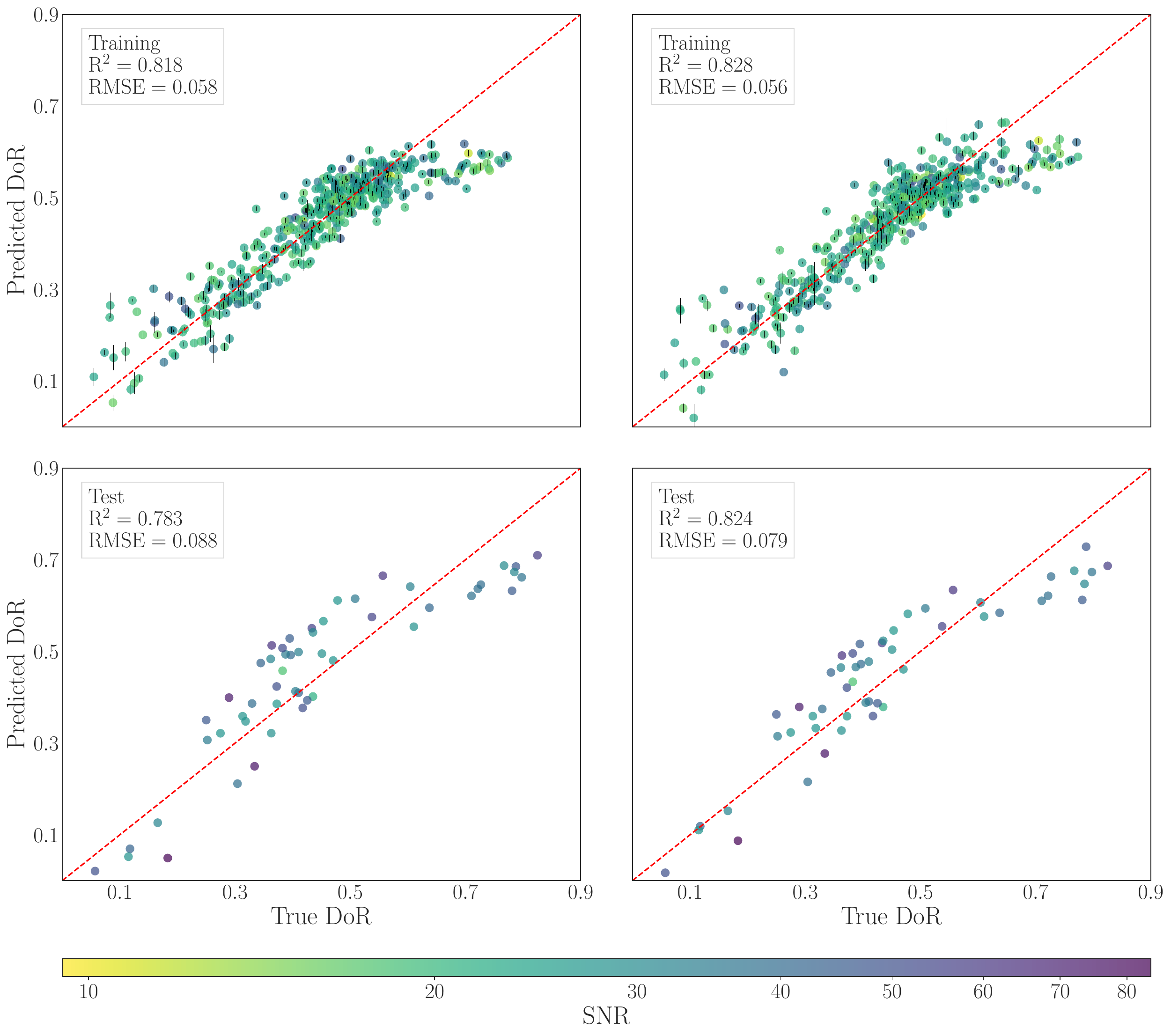}
    \caption{True versus Predicted DoR from the worst (left) and best (right) performing model (on the test data) for the training data (top) and the test data (bottom). Error bars are the standard deviation in prediction across the random seeds used in creating the k-folds. The mean $R^2$ and root-mean-squared (RMS) errors are annotated. Points are coloured according to the galaxy spectrum's SNR. The red dashed line shows the perfect prediction. 
    }
\label{fig:regression_performance}
\end{figure*}

Figure~\ref{fig:regression_performance} shows the behaviour of the regression models at the two extremes of performance on the training (top) and test (bottom) samples. The left panels correspond to the worst performing configuration, while the right panels illustrate the best-performing one. 

We compare the predicted DoR from the SVR algorithm (y-axis) with the 'True DoR', as obtained from the careful spectroscopic stellar population analysis \citepalias{Spiniello24,Mills25}.  The data points are colour-coded according to the SNR of the spectra\footnote{We note that the X-Shooter \INSPIRE\ spectra have much higher SNR than the SDSS \EINSPIRE\ ones}, to demonstrate that no dependency on SNR is found for our predictions.

The bottom panels, which shows the model predictions on the test data, reveal an interesting behaviour that is present, to varying degrees, across all configurations: at the extremes of the DoR distribution (i.e. for DoR $\ge 0.6$ and DoR $\le 0.3$), the SVR systematically underestimates the DoR, whereas in the central region ($0.3 < \mathrm{DoR} < 0.6$) it overestimates it. Interestingly, these values broadly match the thresholds identified manually in \citetalias{Spiniello24} and 
\citetalias{Mills25} and used to separate relics from non-relics and extreme relics from relics\footnote{The latter threshold is slightly higher (0.7) in the manual classificaiton}. These turning points remain stable under a variety of methodological tests, including adjustments for class imbalance, the use of different regression algorithms, ensemble approaches, and alternative data encodings, such as one-hot encoding of the [Mg/Fe] values. One-hot encoding is a standard technique used to represent discrete or categorical values as separate binary variables (0 or 1), allowing the model to treat each category independently rather than assuming a continuous numerical ordering. 

This robustness suggests that the observed structure is not driven by a specific modelling choice, but instead reflects underlying preferred regimes in the DoR distribution. These regimes align with the compact galaxy ``families'' discussed in previous observational and theoretical works (e.g., \citealt{Grebol2023, Moura24}; see also the classifications presented in INSPIRE DR1, \citealt{Spiniello+21} and \citetalias{Spiniello24}). In this sense, the regression framework does not impose discrete classes, but naturally recovers transitions that are consistent with a three-family interpretation. Regression is adopted here to retain sensitivity to intermediate regimes and to investigate feature importance, but we note that alternative strategies may prove advantageous when applied to very large survey datasets.

However, it is important to emphasise that the presence of turning points, where predictive performance degrades at the extremes of the DoR distribution, is primarily driven by the structure of the feature space rather than by sample size alone. In the high-DoR regime, several key input features most notably $\tau_{\rm rel}$ and metallicity, exhibit a plateau behaviour, becoming only weakly sensitive to further increases in DoR (see Figure~\ref{fig:corner_plain}). As a consequence, even a substantially larger number of extreme relics would not fully resolve this limitation: once the observables saturate, the regression model lacks sufficient discriminatory information to distinguish among very high DoR values. Mitigating this behaviour would require additional features that retain sensitivity in the extreme relic regime. 
Data scarcity nevertheless contributes to the effect. As shown in the DoR histogram of Figure~\ref{fig:histogram}, extreme relics are intrinsically rare, and the training sample contains relatively few galaxies at both very high and very low DoR. Moreover, both surveys were designed to preferentially target red systems, further reducing the representation of low-DoR galaxies. This uneven sampling limits the model’s ability to learn detailed structure in these sparsely populated regions.

Finally, modelling assumptions also play a role. In the low-DoR regime, the single stellar population (SSP) approximation becomes less appropriate, as such galaxies may host more extended and complex star-formation histories. In this case, age and metallicity constraints are intrinsically less precise, with larger uncertainties arising from the averaging of multiple template components. Together, feature saturation, limited representation, and modelling limitations contribute to the observed performance degradation at the distribution extremes.

To investigate which features drive the distinction between relics and non-relics, we train an SVC classifier at a range of DoR thresholds ($0.30$--$0.65$) and measure permutation importance at each of them. The threshold defines the binary boundary: galaxies with DoR above the threshold are labelled as relics. This allows us to assess whether the most discriminative features change depending on how strictly a ``relic'' is defined. We do not extend above DoR$=0.65$ because the positive class becomes too small for reliable classification (only $\sim6\%$ of the training sample). 

Feature importance is evaluated using permutation importance. Each feature is randomly shuffled in turn and the resulting decrease in $R^2$ is measured. Features producing larger performance degradation are considered more informative. This approach is model-agnostic and particularly suited to kernel-based methods. Permutation importances are computed independently across cross-validation folds and are found to be stable, with uniformly small variance in the importance ranking. We note that permutation importance can be biased in the presence of correlated features; in particular, effective radius $R_e$ and stellar mass $\log (M_\star/M_{\odot})$ are moderately correlated in the UCMG sample. However, both quantities exhibit negligible importance across all thresholds, and therefore any correlation-induced bias does not affect the main interpretation of the results.

\begin{figure}
    \centering    \includegraphics[width=0.99\linewidth]{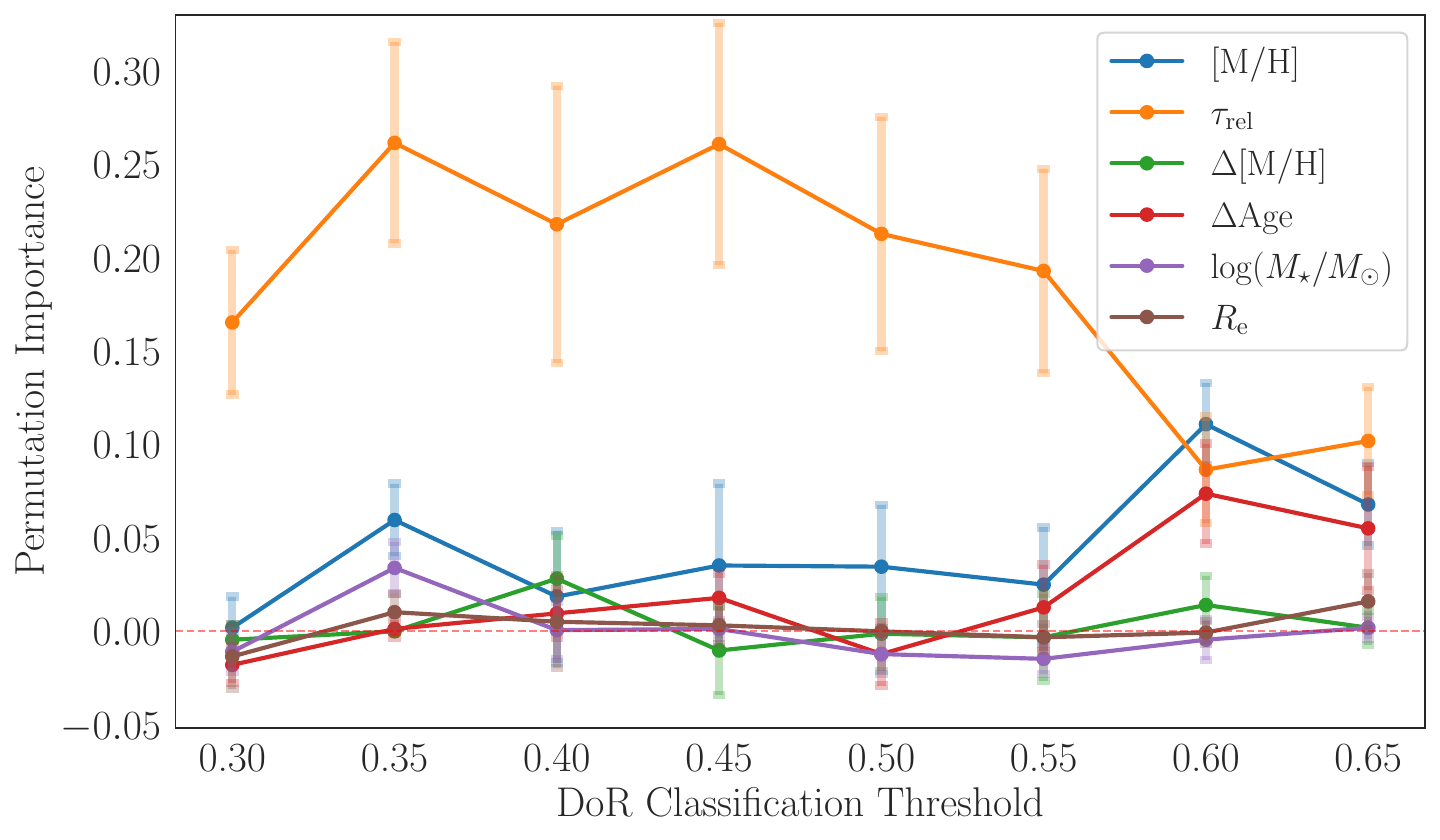}
    \caption{SVC permutation importance for each Euclid-like input feature as a function of the DoR classification threshold. The threshold defines the binary boundary used to label galaxies as relics (DoR $\geq$ threshold) or non-relics. Error bars show the standard deviation across the 30 random feature shuffles used to estimate each importance value.}
    \label{fig:classification}
\end{figure}

The results of this test are shown in Figure~\ref{fig:classification}. Across nearly all thresholds, $\tau_{\rm rel}$ is by far the most important feature, especially at relatively lower DoR thresholds. Given that DoR is derived from star formation histories, its dependence on age-related quantities is expected. At the highest thresholds (DoR$\ge0.6$), however,  the importance of $\tau_{\rm rel}$ decreases as the stellar populations of more extreme relics are uniformly old and $\tau_{\rm rel}$ tends to zero, reducing its discriminating power. In this regime, metallicity becomes a more important feature (importance $\sim0.11$ at DoR$=0.6$), followed by $\Delta\text{age}$ (importance $\sim0.07$). In central DoR regions, although metallicity still correlates well with DoR, it is secondary to $\tau_{\rm rel}$ as there are metal-rich galaxies that are not necessarily relics. Finally, the structural parameters $\log M_{\star}$ and $R_{\rm e}$, as well as $\Delta$[M/H], show negligible or zero importance at all thresholds, confirming that these quantities do not, in this instance, aid in distinguishing relics from non-relics within a sample already selected to be ultra-compact and massive. We stress, however that in this prototype exercise we restricted ourselves to a pre-selected sample of confirmed UCMGs, covering a narrow range of sizes. 
Nevertheless, and quite importantly, for future applications, a cut-off above the upper turning point, i.e. a predicted DoR of $\gtrsim 0.6$, would  allow us to select, with confidence, only the most extreme-relics given that the model tends to systematically under-predict in this region. 

For a binary classifier at a given DoR threshold, we define precision as the fraction of galaxies predicted as relics that are true relics and recall as the fraction of true relics correctly identified by the model. At DoR $\gtrsim 0.6$, the best-performing feature set (model 5) achieves a precision of 0.91 and recall of 0.83 (10/12 extreme relics recovered), while the Euclid-like feature set achieves precision 0.77 with the same recall, reflecting more false positives. Hence, in conclusion, although the testing performance is worse for extreme relics, even the worst performing model remains a useful method in finding these galaxies with an $R^2$ of $0.783$. 

The changing feature importance for different DoR thresholds also suggests that different regions of the DoR space may benefit from different model configurations. For instance, in the high DoR region where more features contribute meaningfully, a more flexible kernel or different hyper-parameters may better capture the complex dependencies. However, given the limited data available, attempts at locally-optimised hyper-parameters across regions of the feature space proved unhelpful.

\section{Discussion and Limitations}
\label{sec:discussion}
This paper, the second of the \EINSPIRE\ project, presents a proof of concept for a machine-learning framework aimed at investigating how the spectroscopically inferred degree of relicness (DoR) of ultra-compact massive galaxies (UCMGs) is encoded within commonly derived stellar population, kinematical, and structural observables. 

Before summarising our main findings, it is important to highlight several methodological considerations and limitations of the current study. First and foremost, the present model is trained on stellar population parameters derived from full spectral fitting, which are measured with significantly higher precision than what will be achievable from Euclid SED-based analyses alone. Although this work demonstrates that DoR can be inferred within a well-characterised spectroscopic framework, its direct applicability to purely photometric datasets will require retraining and validation using survey-level observables. In this sense, the current setup establishes methodological feasibility rather than a fully Euclid-ready implementation.

Second, several of the input features (e.g. age and metallicity) are derived from the same star-formation history reconstruction that defines the DoR itself, although only age-related quantities enter explicitly into the DoR equation \citep{Spiniello24}. The regression therefore operates within a self-consistent stellar population framework and may partly reflect internal correlations of the spectral fitting procedure. This does not invalidate the predictive approach, but implies that the model approximates the spectroscopically inferred DoR rather than an entirely independent observable. In this sense, the present framework should not be interpreted as a replacement for detailed stellar population analyses, but rather as a quantitative tool for investigating how robustly the relic signal manifests across observable parameter space.  Future extensions based on purely photometric or structural quantities will give a more independent validation.

Third, the tendency of the model to underpredict extreme DoR values reflects both the limited size of the training set and the intrinsic behaviour of regression methods, which generally tend to regress toward the mean. While this does not prevent the identification of high-probability relic candidates, it highlights the importance of enlarging the training set and exploring complementary modelling strategies as larger datasets become available. In this respect, the forthcoming \EINSPIRE\ Paper~III, which will analyse UCMGs selected from the DESI Data Release 1 (DR1, \citealt{DESIDR1}), will provide a significantly expanded spectroscopic sample that can be used to retrain and further calibrate the predictive framework. 

Finally, we also note that including parameter uncertainties as input features does not constitute a full propagation of observational errors through the model; a probabilistic framework will be required to quantify predictive uncertainties more rigorously.

Despite these limitations, the framework developed here provides a promising first step toward developing scalable strategies for identifying relic candidates in future large-scale surveys. Such candidate samples may ultimately provide new constraints on the fraction of galaxies that avoided significant size growth after the early assembly phase and allow systematic tests of the relative importance of minor mergers, accretion, and internal dynamical processes in shaping massive galaxy evolution. 
More specifically, the present analysis demonstrates that the empirical relations linking the DoR to observable stellar population and kinematical properties remain stable across independent UCMG datasets, even under moderate covariate shifts, in agreement with the strong empirical correlations between DoR, metallicity, [$\mathrm{Mg}/\mathrm{Fe}$], and stellar velocity dispersion previously identified in \citetalias{Mills25} and \citetalias{Spiniello24}. 

The successful prediction on the independent \INSPIRE\ sample should not be interpreted as evidence that the model is universally applicable to arbitrary galaxy populations. Rather, it demonstrates that the empirical relation linking observable galaxy properties to DoR remains stable across two independently constructed spectroscopic UCMG catalogues despite moderate differences in their underlying distributions. This robustness is further supported by the comparable performance obtained when the two surveys are pooled and randomly repartitioned into training and testing subsets.

\section{Conclusions}
\label{sec:conclusions}
The main results of this study can be summarised as follows.

\begin{itemize}
    \item[(i)] A regression model trained on the relatively small but well-characterised \EINSPIRE\ spectroscopic dataset achieves stable predictive performance when applied to unseen test data from the \INSPIRE\ survey, demonstrating that the DoR can be recovered with good reliability within this framework.
    
    \item[(ii)] The dominant discriminating power lies in age-related quantities, in particular the redshift-normalised age $\tau_{\rm rel}$, which emerges as the most informative feature. The combination of $\tau_{\rm rel}$ and metallicity alone already provides substantial predictive capability, which further improves when their associated uncertainties are included as additional inputs.
    
    \item[(iii)] Additional spectroscopic information, such as [Mg/Fe] and stellar velocity dispersion, enhances the predictive performance but plays a secondary role compared to age-related parameters. Structural quantities (stellar mass and effective radius) primarily define the UCMG selection and contribute less to distinguishing relics within that population, although this may partly reflect the current measurement uncertainties affecting these parameters.
    
    \item[(iv)] The systematic underprediction of DoR at high values has an important practical implication. By selecting galaxies with predicted DoR $\ge 0.6$, we obtain a conservative and high-purity sample of relic candidates. Since the regression model tends to underestimate the most extreme DoR values, objects above this threshold are very unlikely to be false positives. This makes the approach particularly efficient for prioritising the most extreme relic systems for follow-up spectroscopy.
\end{itemize}

Taken together, these results confirm that the spectroscopically inferred DoR is robustly connected to a set of observable stellar population and kinematical properties across independent UCMG datasets, and demonstrate that these empirical relations remain stable and predictive within a machine-learning regression framework. Within the present framework, machine-learning regression therefore provides a useful quantitative tool for investigating the empirical relations linking relicness to observable galaxy properties. 

Our findings indicate that the relic signal is primarily encoded in stellar population properties, reinforcing earlier results from \INSPIRE\ and \EINSPIRE\ that link DoR to metallicity, velocity dispersion, and $\alpha$-enhancement, while clarifying their relative importance within a predictive framework. 
Although the present implementation still relies on spectroscopically derived stellar population quantities, the results suggest that some aspects of the relic signal may remain recoverable from a limited set of observable galaxy properties. In the context of forthcoming wide-area surveys, this framework may therefore help guide future pre-selection strategies for identifying high-probability relic candidates for detailed, high-SNR stellar population analyses enabled by tailored spectroscopic follow-up. Indeed, applied to Euclid-scale datasets, and appropriately retrained on survey-level observables, the method can potentially enable the construction of large, high-purity candidate samples that can be prioritised for spectroscopic follow-up with facilities such as DESI and future high-resolution instruments.
In summary, this study establishes that the empirical relations underlying the DoR formalism remain stable and predictive across independent spectroscopic UCMG samples. More generally, the analysis provides a framework for investigating how relicness manifests across observable parameter space and for guiding the future development of scalable relic-candidate selection strategies in next-generation surveys. 

\section*{Acknowledgements}
The authors wish to thank all collaborators from the \INSPIRE\ and \EINSPIRE\ surveys for constructive and interesting comments.\\  
AFM acknowledges support from RYC2021-031099-I and PID2024-162088NB-I00 of MICIN/AEI/10.13039/501100011033/ UE NextGenerationEU/PRTR. 
J.H. acknowledges support from TCSMT through a starting grant. 
M.S. acknowledges support by the State Research Agency of the Spanish Ministry of Science and Innovation under the grants 'Galaxy Evolution with Artificial Intelligence' (PGC2018-100852-A-I00) and 'BASALT' (PID2021-126838NB-I00) and the Polish National Agency for Academic Exchange (Bekker grant BPN/BEK/2021/1/00298/DEC/1). 


\section*{Data Availability}
The \INSPIRE\ catalogue and spectra are  publicly available through the ESO Phase 3 
Archive Science Portal under the collection \INSPIRE: \url{https://archive.eso.org/scienceportal/home?data_collection=INSPIRE}. The \EINSPIRE\ catalogue is instead publicly available from the survey website: \url{https://sites.google.com/inaf.it/chiara-spiniello/e-inspire}. Single SDSS spectra can be obtained from the PI under specific requests or downloaded directly through the SDSS archive using the coordinates present in the catalogue. 


\bibliographystyle{mnras}
\bibliography{biblio_INSPIRE}

@ARTICLE{Tortora25,
       author = {{Tortora}, C. and {Tozzi}, G. and {Agapito}, G. and {Barbera}, F. La and {Spiniello}, C. and {Li}, R. and {Carl{\`a}}, G. and {D'Ago}, G. and {Ghose}, E. and {Mannucci}, F. and {Napolitano}, N.~R. and {Pinna}, E. and {Arnaboldi}, M. and {Bevacqua}, D. and {Ferr{\'e}-Mateu}, A. and {Gallazzi}, A. and {Hartke}, J. and {Hunt}, L.~K. and {Maksymowicz-Maciata}, M. and {Pulsoni}, C. and {Saracco}, P. and {Scognamiglio}, D. and {Spavone}, M.},
        title = "{INSPIRE: INvestigating Stellar Populations In RElics - IX. KiDS J0842 + 0059: the first fully confirmed relic beyond the local Universe}",
      journal = {\mnras},
         year = 2025,
        month = jul,
       volume = {540},
       number = {3},
        pages = {2555-2565},
          doi = {10.1093/mnras/staf831},
archivePrefix = {arXiv},
       eprint = {2505.13611},
 primaryClass = {astro-ph.GA},
       adsurl = {https://ui.adsabs.harvard.edu/abs/2025MNRAS.540.2555T}
}

@ARTICLE{Abdurrouf25,
       author = {{Euclid Collaboration: Abdurro'uf} and {Tortora}, C. and {Baes}, M. and {Nersesian}, A. and {Kovacic}, I. and {Bolzonella}, M. and {Lancon}, A. and {Bisigello}, L. and {Annibali}, F. and {Bremer}, M.~N. and {Carollo}, D. and {Conselice}, C.~J. and {Enia}, A. and {Ferguson}, A.~M.~N. and {Ferre-Mateu}, A. and {Hunt}, L.~K. and {Iodice}, E. and {Knapen}, J.~H. and {Iovino}, A. and {Marleau}, F.~R. and {Peletier}, R.~F. and {Ragusa}, R. and {Rejkuba}, M. and {Robotham}, A.~S.~G. and {Roman}, J. and {Saifollahi}, T. and {Salucci}, P. and {Scodeggio}, M. and {Siudek}, M. and {van der Wel}, A. and {Voggel}, K. and {Altieri}, B. and {Andreon}, S. and {Baccigalupi}, C. and {Baldi}, M. and {Bardelli}, S. and {Biviano}, A. and {Bonchi}, A. and {Bonino}, D. and {Branchini}, E. and {Brescia}, M. and {Brinchmann}, J. and {Caillat}, A. and {Camera}, S. and {Canas-Herrera}, G. and {Capobianco}, V. and {Carbone}, C. and {Carretero}, J. and {Casas}, S. and {Castellano}, M. and {Castignani}, G. and {Cavuoti}, S. and {Chambers}, K.~C. and {Cimatti}, A. and {Colodro-Conde}, C. and {Congedo}, G. and {Conversi}, L. and {Copin}, Y. and {Courbin}, F. and {Courtois}, H.~M. and {Cropper}, M. and {da Silva}, A. and {Degaudenzi}, H. and {de}, Lucia G. and {di}, Giorgio A.~M. and {Dinis}, J. and {Dole}, H. and {Dubath}, F. and {Dupac}, X. and {Dusini}, S. and {Escoffier}, S. and {Farina}, M. and {Farinelli}, R. and {Farrens}, S. and {Faustini}, F. and {Ferriol}, S. and {Finelli}, F. and {Fotopoulou}, S. and {Frailis}, M. and {Franceschi}, E. and {Fumana}, M. and {Galeotta}, S. and {Gillis}, B. and {Giocoli}, C. and {Gomez-Alvarez}, P. and {Gracia-Carpio}, J. and {Grazian}, A. and {Grupp}, F. and {Holmes}, W. and {Hormuth}, F. and {Hornstrup}, A. and {Hudelot}, P. and {Jahnke}, K. and {Jhabvala}, M. and {Keihanen}, E. and {Kermiche}, S. and {Kiessling}, A. and {Kilbinger}, M. and {Kubik}, B. and {Kummel}, M. and {Kunz}, M. and {Kurki-Suonio}, H. and {Le Brun}, A.~M.~C. and {Ligori}, S. and {Lilje}, P.~B. and {Lindholm}, V. and {Lloro}, I. and {Mainetti}, G. and {Maino}, D. and {Maiorano}, E. and {Mansutti}, O. and {Marggraf}, O. and {Markovic}, K. and {Martinelli}, M. and {Martinet}, N. and {Marulli}, F. and {Massey}, R. and {Medinaceli}, E. and {Mei}, S. and {Melchior}, M. and {Mellier}, Y. and {Meneghetti}, M. and {Merlin}, E. and {Meylan}, G. and {Mora}, A. and {Moresco}, M. and {Moscardini}, L. and {Niemi}, S.-M. and {Nightingale}, J.~W. and {Padilla}, C. and {Paltani}, S. and {Pasian}, F. and {Pedersen}, K. and {Pettorino}, V. and {Polenta}, G. and {Poncet}, M. and {Popa}, L.~A. and {Pozzetti}, L. and {Raison}, F. and {Renzi}, A. and {Rhodes}, J. and {Riccio}, G. and {Romelli}, E. and {Roncarelli}, M. and {Rossetti}, E. and {Saglia}, R. and {Sakr}, Z. and {Sapone}, D. and {Sartoris}, B. and {Schirmer}, M. and {Schneider}, P. and {Schrabback}, T. and {Secroun}, A. and {Sefusatti}, E. and {Seidel}, G. and {Serrano}, S. and {Simon}, P. and {Sirignano}, C. and {Sirri}, G. and {Stanco}, L. and {Steinwagner}, J. and {Tallada-Crespi}, P. and {Taylor}, A.~N. and {Tereno}, I. and {Toft}, S. and {Toledo-Moreo}, R. and {Torradeflot}, F. and {Tutusaus}, I. and {Valenziano}, L. and {Valiviita}, J. and {Vassallo}, T. and {Verdoes Kleijn}, G. and {Veropalumbo}, A. and {Wang}, Y. and {Weller}, J. and {Zamorani}, G. and {Zucca}, E. and {Bozzo}, E. and {Burigana}, C. and {Calabrese}, M. and {di Ferdinando}, D. and {Escartin Vigo}, J.~A. and {Matthew}, S. and {Mauri}, N. and {Pontinen}, M. and {Porciani}, C. and {Scottez}, V. and {Tenti}, M. and {Viel}, M. and {Wiesmann}, M. and {Akrami}, Y. and {Allevato}, V. and {Anselmi}, S. and {Archidiacono}, M. and {Atrio-Barandela}, F. and {Ballardini}, M. and {Bertacca}, D. and {Blanchard}, A. and {Blot}, L.},
        title = "{Euclid preparation: LXXIII. Spatially resolved stellar populations of local galaxies with Euclid: A proof of concept using synthetic images with the TNG50 simulation}",
      journal = {\aap},
         year = 2025,
        month = oct,
       volume = {702},
          eid = {A72},
        pages = {A72},
          doi = {10.1051/0004-6361/202554516},
archivePrefix = {arXiv},
       eprint = {2503.15635},
 primaryClass = {astro-ph.GA},
       adsurl = {https://ui.adsabs.harvard.edu/abs/2025A&A...702A..72E}
}

@ARTICLE{Nersesian25,
       author = {{Euclid Collaboration: Nersesian}, A. and {Abdurro'uf} and {Baes}, M. and {Tortora}, C. and {Kova{\v{c}}i{\'c}}, I. and {Bisigello}, L. and {Corcho-Caballero}, P. and {Dur{\'a}n-Camacho}, E. and {Hunt}, L.~K. and {Iglesias-Navarro}, P. and {Ragusa}, R. and {Rom{\'a}n}, J. and {Shankar}, F. and {Siudek}, M. and {Sorce}, J.~G. and {Marleau}, F.~R. and {Aghanim}, N. and {Andreon}, S. and {Auricchio}, N. and {Baccigalupi}, C. and {Baldi}, M. and {Bardelli}, S. and {Biviano}, A. and {Branchini}, E. and {Brescia}, M. and {Camera}, S. and {Ca{\~n}as-Herrera}, G. and {Capobianco}, V. and {Carbone}, C. and {Carretero}, J. and {Casas}, S. and {Castellano}, M. and {Castignani}, G. and {Cavuoti}, S. and {Cimatti}, A. and {Colodro-Conde}, C. and {Congedo}, G. and {Conselice}, C.~J. and {Conversi}, L. and {Copin}, Y. and {Courbin}, F. and {Courtois}, H.~M. and {Da Silva}, A. and {Degaudenzi}, H. and {De Lucia}, G. and {Dole}, H. and {Douspis}, M. and {Dubath}, F. and {Dupac}, X. and {Dusini}, S. and {Farina}, M. and {Farinelli}, R. and {Faustini}, F. and {Ferriol}, S. and {Finelli}, F. and {Fourmanoit}, N. and {Frailis}, M. and {Franceschi}, E. and {Fumana}, M. and {Galeotta}, S. and {George}, K. and {Gillis}, B. and {Giocoli}, C. and {Gracia-Carpio}, J. and {Grazian}, A. and {Grupp}, F. and {Haugan}, S.~V.~H. and {Holmes}, W. and {Hormuth}, F. and {Hornstrup}, A. and {Jahnke}, K. and {Jhabvala}, M. and {Keih{\"a}nen}, E. and {Kermiche}, S. and {Kilbinger}, M. and {Kubik}, B. and {K{\"u}mmel}, M. and {Kunz}, M. and {Kurki-Suonio}, H. and {Le Brun}, A.~M.~C. and {Ligori}, S. and {Lilje}, P.~B. and {Lindholm}, V. and {Lloro}, I. and {Mainetti}, G. and {Maino}, D. and {Maiorano}, E. and {Mansutti}, O. and {Marggraf}, O. and {Martinelli}, M. and {Martinet}, N. and {Marulli}, F. and {Massey}, R.~J. and {Medinaceli}, E. and {Mei}, S. and {Melchior}, M. and {Mellier}, Y. and {Meneghetti}, M. and {Merlin}, E. and {Meylan}, G. and {Mora}, A. and {Moresco}, M. and {Moscardini}, L. and {Neissner}, C. and {Niemi}, S.-M. and {Padilla}, C. and {Paltani}, S. and {Pasian}, F. and {Pedersen}, K. and {Pettorino}, V. and {Pires}, S. and {Polenta}, G. and {Poncet}, M. and {Popa}, L.~A. and {Pozzetti}, L. and {Renzi}, A. and {Rhodes}, J. and {Riccio}, G. and {Romelli}, E. and {Roncarelli}, M. and {Saglia}, R. and {Sakr}, Z. and {Sapone}, D. and {Sartoris}, B. and {Schneider}, P. and {Schrabback}, T. and {Secroun}, A. and {Seidel}, G. and {Seiffert}, M. and {Serrano}, S. and {Simon}, P. and {Sirignano}, C. and {Sirri}, G. and {Stanco}, L. and {Steinwagner}, J. and {Tallada-Cresp{\'\i}}, P. and {Taylor}, A.~N. and {Tereno}, I. and {Toft}, S. and {Toledo-Moreo}, R. and {Torradeflot}, F. and {Tutusaus}, I. and {Valenziano}, L. and {Valiviita}, J. and {Vassallo}, T. and {Verdoes Kleijn}, G. and {Veropalumbo}, A. and {Wang}, Y. and {Weller}, J. and {Zamorani}, G. and {Zerbi}, F.~M. and {Zinchenko}, I.~A. and {Zucca}, E. and {Allevato}, V. and {Bolzonella}, M. and {Bozzo}, E. and {Burigana}, C. and {Cabanac}, R. and {Calabrese}, M. and {Cappi}, A. and {Escartin Vigo}, J.~A. and {Gabarra}, L. and {Mart{\'\i}n-Fleitas}, J. and {Matthew}, S. and {Mauri}, N. and {Metcalf}, R.~B. and {Nucita}, A.~A. and {Pezzotta}, A. and {P{\"o}ntinen}, M. and {Porciani}, C. and {Risso}, I. and {Scottez}, V. and {Sereno}, M. and {Tenti}, M. and {Viel}, M. and {Wiesmann}, M. and {Akrami}, Y. and {Andika}, I.~T. and {Anselmi}, S. and {Archidiacono}, M. and {Atrio-Barandela}, F. and {Bertacca}, D. and {Bethermin}, M. and {Blanchard}, A. and {Blot}, L. and {Borgani}, S. and {Brown}, M.~L. and {Bruton}, S. and {Calabro}, A. and {Camacho Quevedo}, B. and {Caro}, F. and {Carvalho}, C.~S. and {Castro}, T. and {Cogato}, F. and {Conseil}, S. and {Cooray}, A.~R. and {Cucciati}, O. and {Davini}, S. and {De Paolis}, F.},
        title = "{Euclid preparation: LXXXI. The impact of nonparametric star formation histories on spatially resolved galaxy property estimation using synthetic Euclid images}",
      journal = {arXiv e-prints},
         year = 2025,
        month = nov,
          eid = {arXiv:2511.22399},
        pages = {arXiv:2511.22399},
          doi = {10.48550/arXiv.2511.22399},
archivePrefix = {arXiv},
       eprint = {2511.22399},
 primaryClass = {astro-ph.GA},
       adsurl = {https://ui.adsabs.harvard.edu/abs/2025arXiv251122399E}
}

@ARTICLE{Kovacic25,
       author = {{Euclid Collaboration: Kovacic}, I. and {Baes}, M. and {Nersesian}, A. and {Andreadis}, N. and {Nemani}, L. and {Abdurro'Uf} and {Bisigello}, L. and {Bolzonella}, M. and {Tortora}, C. and {van der Wel}, A. and {Cavuoti}, S. and {Conselice}, C.~J. and {Enia}, A. and {Hunt}, L.~K. and {Iglesias-Navarro}, P. and {Iodice}, E. and {Knapen}, J.~H. and {Marleau}, F.~R. and {Muller}, O. and {Peletier}, R.~F. and {Roman}, J. and {Ragusa}, R. and {Salucci}, P. and {Saifollahi}, T. and {Scodeggio}, M. and {Siudek}, M. and {de Waele}, T. and {Amara}, A. and {Andreon}, S. and {Auricchio}, N. and {Baccigalupi}, C. and {Baldi}, M. and {Bardelli}, S. and {Battaglia}, P. and {Bender}, R. and {Bodendorf}, C. and {Bonino}, D. and {Bon}, W. and {Branchini}, E. and {Brescia}, M. and {Brinchmann}, J. and {Camera}, S. and {Capobianco}, V. and {Carbone}, C. and {Carretero}, J. and {Casas}, S. and {Castander}, F.~J. and {Castellano}, M. and {Castignani}, G. and {Cimatti}, A. and {Colodro-Conde}, C. and {Congedo}, G. and {Conversi}, L. and {Copin}, Y. and {Courbin}, F. and {Courtois}, H.~M. and {da Silva}, A. and {Degaudenzi}, H. and {de}, Lucia G. and {di}, Giorgio A.~M. and {Dinis}, J. and {Douspis}, M. and {Dubath}, F. and {Dupac}, X. and {Dusini}, S. and {Ealet}, A. and {Farina}, M. and {Farrens}, S. and {Faustini}, F. and {Ferriol}, S. and {Fosalba}, P. and {Frailis}, M. and {Franceschi}, E. and {Galeotta}, S. and {Gillis}, B. and {Giocoli}, C. and {Grazian}, A. and {Grupp}, F. and {Guzzo}, L. and {Haugan}, S.~V.~H. and {Holmes}, W. and {Hook}, I. and {Hormuth}, F. and {Hornstrup}, A. and {Jahnke}, K. and {Jhabvala}, M. and {Joachimi}, B. and {Keihanen}, E. and {Kermiche}, S. and {Kiessling}, A. and {Kilbinger}, M. and {Kubik}, B. and {Kuijken}, K. and {Kummel}, M. and {Kunz}, M. and {Kurki-Suonio}, H. and {Ligori}, S. and {Lilje}, P.~B. and {Lindholm}, V. and {Lloro}, I. and {Maino}, D. and {Maiorano}, E. and {Mansutti}, O. and {Marcin}, S. and {Marggraf}, O. and {Markovic}, K. and {Martinelli}, M. and {Martinet}, N. and {Marulli}, F. and {Massey}, R. and {Medinaceli}, E. and {Mei}, S. and {Melchior}, M. and {Mellier}, Y. and {Meneghetti}, M. and {Merlin}, E. and {Meylan}, G. and {Moresco}, M. and {Moscardini}, L. and {Niemi}, S.-M. and {Nightingale}, J.~W. and {Padilla}, C. and {Paltani}, S. and {Pasian}, F. and {Pedersen}, K. and {Pettorino}, V. and {Pires}, S. and {Polenta}, G. and {Poncet}, M. and {Popa}, L.~A. and {Pozzetti}, L. and {Raison}, F. and {Rebolo}, R. and {Renzi}, A. and {Rhodes}, J. and {Riccio}, G. and {Romelli}, E. and {Roncarelli}, M. and {Rossetti}, E. and {Saglia}, R. and {Sakr}, Z. and {Sanchez}, A.~G. and {Sapone}, D. and {Sartoris}, B. and {Schirmer}, M. and {Schneider}, P. and {Schrabback}, T. and {Secroun}, A. and {Seidel}, G. and {Serrano}, S. and {Sirignano}, C. and {Sirri}, G. and {Stanco}, L. and {Steinwagner}, J. and {Tallada-Crespi}, P. and {Tavagnacco}, D. and {Taylor}, A.~N. and {Teplitz}, H.~I. and {Tereno}, I. and {Toledo-Moreo}, R. and {Torradeflot}, F. and {Tutusaus}, I. and {Valenziano}, L. and {Vassallo}, T. and {Verdoes Kleijn}, G. and {Veropalumbo}, A. and {Wang}, Y. and {Weller}, J. and {Zacchei}, A. and {Zamorani}, G. and {Zucca}, E. and {Biviano}, A. and {Bozzo}, E. and {Burigana}, C. and {Calabrese}, M. and {di Ferdinando}, D. and {Escartin Vigo}, J.~A. and {Finelli}, F. and {Gracia-Carpio}, J. and {Matthew}, S. and {Mauri}, N. and {Pontinen}, M. and {Scottez}, V. and {Tenti}, M. and {Viel}, M. and {Wiesmann}, M. and {Akrami}, Y. and {Allevato}, V. and {Alvi}, S. and {Anselmi}, S. and {Archidiacono}, M. and {Atrio-Barandela}, F. and {Ballardini}, M. and {Bethermin}, M. and {Blot}, L. and {Borgani}, S. and {Bruton}, S. and {Cabanac}, R. and {Calabro}, A.},
        title = "{Euclid preparation: LXVIII. Extracting physical parameters from galaxies with machine learning}",
      journal = {\aap},
         year = 2025,
        month = mar,
       volume = {695},
          eid = {A284},
        pages = {A284},
          doi = {10.1051/0004-6361/202453111},
archivePrefix = {arXiv},
       eprint = {2501.14408},
 primaryClass = {astro-ph.GA},
       adsurl = {https://ui.adsabs.harvard.edu/abs/2025A&A...695A.284E}
}

@ARTICLE{Moura24,
       author = {{Moura}, Micheli T. and {Chies-Santos}, Ana L. and {Furlanetto}, Cristina and {Zhu}, Ling and {Canossa-Gosteinski}, Marco A.},
        title = "{The internal dynamics and environments of Relics and compact massive ETGs with TNG50}",
      journal = {\mnras},
         year = 2024,
        month = feb,
       volume = {528},
       number = {1},
        pages = {353-364},
          doi = {10.1093/mnras/stae013},
archivePrefix = {arXiv},
       eprint = {2401.02798},
 primaryClass = {astro-ph.GA},
       adsurl = {https://ui.adsabs.harvard.edu/abs/2024MNRAS.528..353M}
}

@ARTICLE{Chen16,
       author = {{Chen}, Tianqi and {Guestrin}, Carlos},
        title = "{XGBoost: A Scalable Tree Boosting System}",
      journal = {arXiv e-prints},
         year = 2016,
        month = mar,
          eid = {arXiv:1603.02754},
        pages = {arXiv:1603.02754},
          doi = {10.48550/arXiv.1603.02754},
archivePrefix = {arXiv},
       eprint = {1603.02754},
 primaryClass = {cs.LG},
       adsurl = {https://ui.adsabs.harvard.edu/abs/2016arXiv160302754C}
}

@inproceedings{drucker1997,
  author    = {Harris Drucker and Christopher J. C. Burges and Linda Kaufman and Alex J. Smola and Vladimir Vapnik},
  title     = {Support Vector Regression Machines},
  booktitle = {Advances in Neural Information Processing Systems 9 (NIPS 1996)},
  editor    = {M. Mozer and M. Jordan and T. Petsche},
  pages     = {155--161},
  year      = {1997},
  publisher = {MIT Press}
}

@ARTICLE{DESIDR1,
       author = {{DESI Collaboration} and {Abdul-Karim}, M. and {Adame}, A.~G. and {Aguado}, D. and {Aguilar}, J. and {Ahlen}, S. and {Alam}, S. and {Aldering}, G. and {Alexander}, D.~M. and {Alfarsy}, R. and {Allen}, L. and {Allende Prieto}, C. and {Alves}, O. and {Anand}, A. and {Andrade}, U. and {Armengaud}, E. and {Avila}, S. and {Aviles}, A. and {Awan}, H. and {Bailey}, S. and {Baleato Lizancos}, A. and {Ballester}, O. and {Bault}, A. and {Bautista}, J. and {BenZvi}, S. and {Beraldo e Silva}, L. and {Bermejo-Climent}, J.~R. and {Beutler}, F. and {Bianchi}, D. and {Blake}, C. and {Blum}, R. and {Bolton}, A.~S. and {Bonici}, M. and {Brieden}, S. and {Brodzeller}, A. and {Brooks}, D. and {Buckley-Geer}, E. and {Burtin}, E. and {Canning}, R. and {Carnero Rosell}, A. and {Carr}, A. and {Carrilho}, P. and {Casas}, L. and {Castander}, F.~J. and {Cereskaite}, R. and {Cervantes-Cota}, J.~L. and {Chaussidon}, E. and {Chaves-Montero}, J. and {Chen}, S. and {Chen}, X. and {Claybaugh}, T. and {Cole}, S. and {Cooper}, A.~P. and {Cousinou}, M. -C. and {Cuceu}, A. and {Davis}, T.~M. and {Dawson}, K.~S. and {de Belsunce}, R. and {de la Cruz}, R. and {de la Macorra}, A. and {de Mattia}, A. and {Deiosso}, N. and {Della Costa}, J. and {Demina}, R. and {Demirbozan}, U. and {DeRose}, J. and {Dey}, A. and {Dey}, B. and {Ding}, J. and {Ding}, Z. and {Doel}, P. and {Douglass}, K. and {Dowicz}, M. and {Ebina}, H. and {Edelstein}, J. and {Eisenstein}, D.~J. and {Elbers}, W. and {Emas}, N. and {Escoffier}, S. and {Fagrelius}, P. and {Fan}, X. and {Fanning}, K. and {Fawcett}, V.~A. and {Fern\textbackslash'andez-Garc\textbackslash'ia}, E. and {Ferraro}, S. and {Findlay}, N. and {Font-Ribera}, A. and {Forero-Romero}, J.~E. and {Forero-S\textbackslash'anchez}, D. and {Frenk}, C.~S. and {G\textbackslash''ansicke}, B.~T. and {Galbany}, L. and {Garc\textbackslash'ia-Bellido}, J. and {Garcia-Quintero}, C. and {Garrison}, L.~H. and {Gazta\textbackslash\raisebox{-0.5ex}\textasciitildenaga}, E. and {Gil-Mar\textbackslash'in}, H. and {Gnedin}, O.~Y. and {Gontcho}, S. Gontcho A and {Gonzalez-Morales}, A.~X. and {Gonzalez-Perez}, V. and {Gordon}, C. and {Graur}, O. and {Green}, D. and {Gruen}, D. and {Gsponer}, R. and {Guandalin}, C. and {Gutierrez}, G. and {Guy}, J. and {Hahn}, C. and {Han}, J.~J. and {Han}, J. and {He}, S. and {Herrera-Alcantar}, H.~K. and {Honscheid}, K. and {Hou}, J. and {Howlett}, C. and {Huterer}, D. and {Ir\textbackslashv\{s\}i\textbackslashv\{c\}}, V. and {Ishak}, M. and {Jacques}, A. and {Jimenez}, J. and {Jing}, Y.~P. and {Joachimi}, B. and {Joudaki}, S. and {Joyce}, R. and {Jullo}, E. and {Juneau}, S. and {Kara\textbackslashc\{c\}ayl\{\textbackslashi\}}, N.~G. and {Karim}, T. and {Kehoe}, R. and {Kent}, S. and {Khederlarian}, A. and {Kirkby}, D. and {Kisner}, T. and {Kitaura}, F. -S. and {Kizhuprakkat}, N. and {Kong}, H. and {Koposov}, S.~E. and {Kremin}, A. and {Krolewski}, A. and {Lahav}, O. and {Lai}, Y. and {Lamman}, C. and {Lan}, T. -W. and {Landriau}, M. and {Lang}, D. and {Lange}, J.~U. and {Lasker}, J. and {Le Goff}, J.~M. and {Le Guillou}, L. and {Leauthaud}, A. and {Levi}, M.~E. and {Li}, S. and {Li}, T.~S. and {Lodha}, K. and {Lokken}, M. and {Luo}, Y. and {Magneville}, C. and {Manera}, M. and {Manser}, C.~J. and {Margala}, D. and {Martini}, P. and {Maus}, M. and {McCullough}, J. and {McDonald}, P. and {Medina}, G.~E. and {Medina-Varela}, L. and {Meisner}, A. and {Mena-Fern\textbackslash'andez}, J. and {Menegas}, A. and {Mezcua}, M. and {Miquel}, R. and {Montero-Camacho}, P. and {Moon}, J. and {Moustakas}, J. and {Mu\textbackslash\raisebox{-0.5ex}\textasciitildenoz-Guti\textbackslash'errez}, A. and {Mu\textbackslash\raisebox{-0.5ex}\textasciitildenoz-Santos}, D. and {Myers}, A.~D. and {Myles}, J. and {Nadathur}, S. and {Najita}, J. and {Napolitano}, L. and {Newman}, J.~A. and {Nikakhtar}, F. and {Nikutta}, R. and {Niz}, G. and {Noriega}, H.~E. and {Padmanabhan}, N. and {Paillas}, E. and {Palanque-Delabrouille}, N. and {Palmese}, A. and {Pan}, J. and {Pan}, Z. and {Parkinson}, D. and {Peacock}, J. and {Percival}, W.~J. and {P\textbackslash'erez-Fern\textbackslash'andez}, A. and {P\textbackslash'erez-R\textbackslash`afols}, I. and {Peterson}, P.},
        title = "{Data Release 1 of the Dark Energy Spectroscopic Instrument}",
      journal = {arXiv e-prints},
         year = 2025,
        month = mar,
          eid = {arXiv:2503.14745},
        pages = {arXiv:2503.14745},
          doi = {10.48550/arXiv.2503.14745},
archivePrefix = {arXiv},
       eprint = {2503.14745},
 primaryClass = {astro-ph.CO},
       adsurl = {https://ui.adsabs.harvard.edu/abs/2025arXiv250314745D}
}

@article{Pedregosa11,
  author  = {Fabian Pedregosa and Ga{{\"e}}l Varoquaux and Alexandre Gramfort and Vincent Michel and Bertrand Thirion and Olivier Grisel and Mathieu Blondel and Peter Prettenhofer and Ron Weiss and Vincent Dubourg and Jake Vanderplas and Alexandre Passos and David Cournapeau and Matthieu Brucher and Matthieu Perrot and {{\'E}}douard Duchesnay},
  title   = {Scikit-learn: Machine Learning in Python},
  journal = {Journal of Machine Learning Research},
  year    = {2011},
  volume  = {12},
  number  = {85},
  pages   = {2825-2830},
  url     = {http://jmlr.org/papers/v12/pedregosa11a.html}
}

@article{Hoerl1970,
author = {Arthur E. Hoerl and Robert W. Kennard and},
title = {Ridge Regression: Biased Estimation for Nonorthogonal Problems},
journal = {Technometrics},
volume = {12},
number = {1},
pages = {55--67},
year = {1970},
publisher = {ASA Website},
doi = {10.1080/00401706.1970.10488634},
URL ={https://www.tandfonline.com/doi/abs/10.1080/00401706.1970.10488634},
eprint = {https://www.tandfonline.com/doi/pdf/10.1080/00401706.1970.10488634}
}

@ARTICLE{Enia25,
       author = {{Euclid Collaboration: Enia}, A. and {Pozzetti}, L. and {Bolzonella}, M. and {Bisigello}, L. and {Hartley}, W.~G. and {Saulder}, C. and {Daddi}, E. and {Siudek}, M. and {Zamorani}, G. and {Cassata}, P. and {Gentile}, F. and {Wang}, L. and {Rodighiero}, G. and {Allevato}, V. and {Corcho-Caballero}, P. and {Dom{\'\i}nguez S{\'a}nchez}, H. and {Tortora}, C. and {Baes}, M. and {Abdurro'uf} and {Nersesian}, A. and {Spinoglio}, L. and {Schaye}, J. and {Ascasibar}, Y. and {Scott}, D. and {Duran-Camacho}, E. and {Quai}, S. and {Talia}, M. and {Mao}, Z. and {Aghanim}, N. and {Altieri}, B. and {Amara}, A. and {Andreon}, S. and {Auricchio}, N. and {Aussel}, H. and {Baccigalupi}, C. and {Baldi}, M. and {Balestra}, A. and {Bardelli}, S. and {Basset}, A. and {Battaglia}, P. and {Bender}, R. and {Biviano}, A. and {Bonchi}, A. and {Branchini}, E. and {Brescia}, M. and {Brinchmann}, J. and {Camera}, S. and {Ca{\~n}as-Herrera}, G. and {Capobianco}, V. and {Carbone}, C. and {Carretero}, J. and {Casas}, S. and {Castander}, F.~J. and {Castellano}, M. and {Castignani}, G. and {Cavuoti}, S. and {Chambers}, K.~C. and {Cimatti}, A. and {Colodro-Conde}, C. and {Congedo}, G. and {Conselice}, C.~J. and {Conversi}, L. and {Copin}, Y. and {Courbin}, F. and {Courtois}, H.~M. and {Cropper}, M. and {Da Silva}, A. and {Degaudenzi}, H. and {De Lucia}, G. and {Di Giorgio}, A.~M. and {Dolding}, C. and {Dole}, H. and {Dubath}, F. and {Duncan}, C.~A.~J. and {Dupac}, X. and {Dusini}, S. and {Ealet}, A. and {Escoffier}, S. and {Fabricius}, M. and {Farina}, M. and {Farinelli}, R. and {Faustini}, F. and {Ferriol}, S. and {Finelli}, F. and {Fotopoulou}, S. and {Frailis}, M. and {Franceschi}, E. and {Franzetti}, P. and {Galeotta}, S. and {George}, K. and {Gillis}, B. and {Giocoli}, C. and {G{\'o}mez-Alvarez}, P. and {Gracia-Carpio}, J. and {Granett}, B.~R. and {Grazian}, A. and {Grupp}, F. and {Guzzo}, L. and {Gwyn}, S. and {Haugan}, S.~V.~H. and {Hoar}, J. and {Holmes}, W. and {Hook}, I.~M. and {Hormuth}, F. and {Hornstrup}, A. and {Hudelot}, P. and {Jahnke}, K. and {Jhabvala}, M. and {Joachimi}, B. and {Keih{\"a}nen}, E. and {Kermiche}, S. and {Kiessling}, A. and {Kubik}, B. and {K{\"u}mmel}, M. and {Kunz}, M. and {Kurki-Suonio}, H. and {Le Boulc'h}, Q. and {Le Brun}, A.~M.~C. and {Le Mignant}, D. and {Ligori}, S. and {Lilje}, P.~B. and {Lindholm}, V. and {Lloro}, I. and {Mainetti}, G. and {Maino}, D. and {Maiorano}, E. and {Mansutti}, O. and {Marggraf}, O. and {Martinelli}, M. and {Martinet}, N. and {Marulli}, F. and {Massey}, R. and {Masters}, D.~C. and {Maurogordato}, S. and {Medinaceli}, E. and {Mei}, S. and {Melchior}, M. and {Mellier}, Y. and {Meneghetti}, M. and {Merlin}, E. and {Meylan}, G. and {Mora}, A. and {Moresco}, M. and {Moscardini}, L. and {Nakajima}, R. and {Neissner}, C. and {Niemi}, S. -M. and {Nightingale}, J.~W. and {Padilla}, C. and {Paltani}, S. and {Pasian}, F. and {Pedersen}, K. and {Percival}, W.~J. and {Pettorino}, V. and {Pires}, S. and {Polenta}, G. and {Poncet}, M. and {Popa}, L.~A. and {Raison}, F. and {Rebolo}, R. and {Renzi}, A. and {Rhodes}, J. and {Riccio}, G. and {Romelli}, E. and {Roncarelli}, M. and {Rossetti}, E. and {Rusholme}, B. and {Saglia}, R. and {Sakr}, Z. and {S{\'a}nchez}, A.~G. and {Sapone}, D. and {Sartoris}, B. and {Schewtschenko}, J.~A. and {Schirmer}, M. and {Schneider}, P. and {Schrabback}, T. and {Scodeggio}, M. and {Secroun}, A. and {Seidel}, G. and {Serrano}, S. and {Simon}, P. and {Sirignano}, C. and {Sirri}, G. and {Skottfelt}, J. and {Stanco}, L. and {Steinwagner}, J. and {Surace}, C. and {Tallada-Cresp{\'\i}}, P. and {Taylor}, A.~N. and {Teplitz}, H.~I. and {Tereno}, I. and {Toft}, S. and {Toledo-Moreo}, R. and {Torradeflot}, F. and {Tutusaus}, I. and {Valenziano}, L. and {Valiviita}, J. and {Vassallo}, T. and {Verdoes Kleijn}, G.},
        title = "{Euclid Quick Data Release (Q1). A first view of the star-forming main sequence in the Euclid Deep Fields}",
      journal = {arXiv e-prints},
         year = 2025,
        month = mar,
          eid = {arXiv:2503.15314},
        pages = {arXiv:2503.15314},
          doi = {10.48550/arXiv.2503.15314},
archivePrefix = {arXiv},
       eprint = {2503.15314},
 primaryClass = {astro-ph.GA},
       adsurl = {https://ui.adsabs.harvard.edu/abs/2025arXiv250315314E}
}

@ARTICLE{Enia24,
       author = {{Euclid Collaboration: Enia} and {Enia}, A. and {Bolzonella}, M. and {Pozzetti}, L. and {Humphrey}, A. and {Cunha}, P.~A.~C. and {Hartley}, W.~G. and {Dubath}, F. and {Paltani}, S. and {Lopez Lopez}, X. and {Quai}, S. and {Bardelli}, S. and {Bisigello}, L. and {Cavuoti}, S. and {De Lucia}, G. and {Ginolfi}, M. and {Grazian}, A. and {Siudek}, M. and {Tortora}, C. and {Zamorani}, G. and {Aghanim}, N. and {Altieri}, B. and {Amara}, A. and {Andreon}, S. and {Auricchio}, N. and {Baccigalupi}, C. and {Baldi}, M. and {Bender}, R. and {Bodendorf}, C. and {Bonino}, D. and {Branchini}, E. and {Brescia}, M. and {Brinchmann}, J. and {Camera}, S. and {Capobianco}, V. and {Carbone}, C. and {Carretero}, J. and {Casas}, S. and {Castander}, F.~J. and {Castellano}, M. and {Castignani}, G. and {Cimatti}, A. and {Colodro-Conde}, C. and {Congedo}, G. and {Conselice}, C.~J. and {Conversi}, L. and {Copin}, Y. and {Corcione}, L. and {Courbin}, F. and {Courtois}, H.~M. and {Da Silva}, A. and {Degaudenzi}, H. and {Di Giorgio}, A.~M. and {Dinis}, J. and {Dupac}, X. and {Dusini}, S. and {Fabricius}, M. and {Farina}, M. and {Farrens}, S. and {Ferriol}, S. and {Fosalba}, P. and {Fotopoulou}, S. and {Frailis}, M. and {Franceschi}, E. and {Fumana}, M. and {Galeotta}, S. and {Gillis}, B. and {Giocoli}, C. and {Grupp}, F. and {Haugan}, S.~V.~H. and {Holmes}, W. and {Hook}, I. and {Hormuth}, F. and {Hornstrup}, A. and {Jahnke}, K. and {Joachimi}, B. and {Keih{\"a}nen}, E. and {Kermiche}, S. and {Kiessling}, A. and {Kubik}, B. and {K{\"u}mmel}, M. and {Kunz}, M. and {Kurki-Suonio}, H. and {Ligori}, S. and {Lilje}, P.~B. and {Lindholm}, V. and {Lloro}, I. and {Maiorano}, E. and {Mansutti}, O. and {Marggraf}, O. and {Markovic}, K. and {Martinelli}, M. and {Martinet}, N. and {Marulli}, F. and {Massey}, R. and {McCracken}, H.~J. and {Medinaceli}, E. and {Mei}, S. and {Melchior}, M. and {Mellier}, Y. and {Meneghetti}, M. and {Merlin}, E. and {Meylan}, G. and {Moresco}, M. and {Moscardini}, L. and {Munari}, E. and {Neissner}, C. and {Niemi}, S. -M. and {Nightingale}, J.~W. and {Padilla}, C. and {Pasian}, F. and {Pedersen}, K. and {Pettorino}, V. and {Polenta}, G. and {Poncet}, M. and {Popa}, L.~A. and {Raison}, F. and {Rebolo}, R. and {Renzi}, A. and {Rhodes}, J. and {Riccio}, G. and {Romelli}, E. and {Roncarelli}, M. and {Rossetti}, E. and {Saglia}, R. and {Sakr}, Z. and {Sapone}, D. and {Schneider}, P. and {Schrabback}, T. and {Scodeggio}, M. and {Secroun}, A. and {Sefusatti}, E. and {Seidel}, G. and {Serrano}, S. and {Sirignano}, C. and {Sirri}, G. and {Stanco}, L. and {Steinwagner}, J. and {Surace}, C. and {Tallada-Cresp{\'\i}}, P. and {Tavagnacco}, D. and {Taylor}, A.~N. and {Teplitz}, H.~I. and {Tereno}, I. and {Toledo-Moreo}, R. and {Torradeflot}, F. and {Tutusaus}, I. and {Valenziano}, L. and {Vassallo}, T. and {Verdoes Kleijn}, G. and {Veropalumbo}, A. and {Wang}, Y. and {Weller}, J. and {Zucca}, E. and {Biviano}, A. and {Boucaud}, A. and {Burigana}, C. and {Calabrese}, M. and {Escartin Vigo}, J.~A. and {Gracia-Carpio}, J. and {Mauri}, N. and {Pezzotta}, A. and {P{\"o}ntinen}, M. and {Porciani}, C. and {Scottez}, V. and {Tenti}, M. and {Viel}, M. and {Wiesmann}, M. and {Akrami}, Y. and {Allevato}, V. and {Anselmi}, S. and {Ballardini}, M. and {Bergamini}, P. and {Bethermin}, M. and {Blanchard}, A. and {Blot}, L. and {Borgani}, S. and {Bruton}, S. and {Cabanac}, R. and {Calabro}, A. and {Canas-Herrera}, G. and {Cappi}, A. and {Carvalho}, C.~S. and {Castro}, T. and {Chambers}, K.~C. and {Contarini}, S. and {Contini}, T. and {Cooray}, A.~R. and {Cucciati}, O. and {Davini}, S. and {De Caro}, B. and {Desprez}, G. and {D{\'\i}az-S{\'a}nchez}, A. and {Di Domizio}, S. and {Dole}, H. and {Escoffier}, S. and {Ferrari}, A.~G. and {Ferreira}, P.~G. and {Ferrero}, I. and {Finoguenov}, A.},
        title = "{Euclid preparation: LI. Forecasting the recovery of galaxy physical properties and their relations with template-fitting and machine-learning methods}",
      journal = {\aap},
         year = 2024,
        month = nov,
       volume = {691},
          eid = {A175},
        pages = {A175},
          doi = {10.1051/0004-6361/202451425},
archivePrefix = {arXiv},
       eprint = {2407.07940},
 primaryClass = {astro-ph.GA},
       adsurl = {https://ui.adsabs.harvard.edu/abs/2024A&A...691A.175E}
}

@ARTICLE{Jones24,
       author = {{Jones}, Evan and {Do}, Tuan and {Boscoe}, Bernie and {Singal}, Jack and {Wan}, Yujie and {Nguyen}, Zooey},
        title = "{Improving Photometric Redshift Estimation for Cosmology with LSST Using Bayesian Neural Networks}",
      journal = {\apj},
         year = 2024,
        month = apr,
       volume = {964},
       number = {2},
          eid = {130},
        pages = {130},
          doi = {10.3847/1538-4357/ad2070},
archivePrefix = {arXiv},
       eprint = {2306.13179},
 primaryClass = {astro-ph.CO},
       adsurl = {https://ui.adsabs.harvard.edu/abs/2024ApJ...964..130J}
}

@ARTICLE{Sharma20,
       author = {{Sharma}, Kaushal and {Kembhavi}, Ajit and {Kembhavi}, Aniruddha and {Sivarani}, T. and {Abraham}, Sheelu and {Vaghmare}, Kaustubh},
        title = "{Application of convolutional neural networks for stellar spectral classification}",
      journal = {\mnras},
         year = 2020,
        month = jan,
       volume = {491},
       number = {2},
        pages = {2280-2300},
          doi = {10.1093/mnras/stz3100},
archivePrefix = {arXiv},
       eprint = {1909.05459},
 primaryClass = {astro-ph.SR},
       adsurl = {https://ui.adsabs.harvard.edu/abs/2020MNRAS.491.2280S}
}

@ARTICLE{Zhao25,
       author = {{Zhao}, Dingyi and {Peng}, Yingjie and {Jing}, Yipeng and {Yang}, Xiaohu and {Ho}, Luis C. and {Renzini}, Alvio and {Gallazzi}, Anna R. and {Lyu}, Cheqiu and {Maiolino}, Roberto and {Dou}, Jing and {Gao}, Zeyu and {Gu}, Qiusheng and {Mannucci}, Filippo and {Mo}, Houjun and {Wang}, Bitao and {Wang}, Enci and {Wang}, Kai and {Wang}, Yu-Chen and {Xu}, Bingxiao and {Yuan}, Feng and {Zhu}, Xingye},
        title = "{From Halos to Galaxies. VI. Improved Halo Mass Estimation for SDSS Groups and Measurement of the Halo Mass Function}",
      journal = {\apj},
         year = 2025,
        month = jan,
       volume = {979},
       number = {1},
          eid = {42},
        pages = {42},
          doi = {10.3847/1538-4357/ad991f},
archivePrefix = {arXiv},
       eprint = {2408.12442},
 primaryClass = {astro-ph.GA},
       adsurl = {https://ui.adsabs.harvard.edu/abs/2025ApJ...979...42Z}
}

@ARTICLE{Humphrey22,
       author = {{Humphrey}, A. and {Kuberski}, W. and {Bialek}, J. and {Perrakis}, N. and {Cools}, W. and {Nuyttens}, N. and {Elakhrass}, H. and {Cunha}, P.~A.~C.},
        title = "{Machine-learning classification of astronomical sources: estimating F1-score in the absence of ground truth}",
      journal = {\mnras},
         year = 2022,
        month = nov,
       volume = {517},
       number = {1},
        pages = {L116-L120},
          doi = {10.1093/mnrasl/slac120},
archivePrefix = {arXiv},
       eprint = {2209.15112},
 primaryClass = {astro-ph.IM},
       adsurl = {https://ui.adsabs.harvard.edu/abs/2022MNRAS.517L.116H}
}

@ARTICLE{Huertas23,
       author = {{Huertas-Company}, M. and {Lanusse}, F.},
        title = "{The Dawes Review 10: The impact of deep learning for the analysis of galaxy surveys}",
      journal = {\pasa},
         year = 2023,
        month = jan,
       volume = {40},
          eid = {e001},
        pages = {e001},
          doi = {10.1017/pasa.2022.55},
archivePrefix = {arXiv},
       eprint = {2210.01813},
 primaryClass = {astro-ph.IM},
       adsurl = {https://ui.adsabs.harvard.edu/abs/2023PASA...40....1H}
}

@ARTICLE{Martin20,
       author = {{Martin}, G. and {Kaviraj}, S. and {Hocking}, A. and {Read}, S.~C. and {Geach}, J.~E.},
        title = "{Galaxy morphological classification in deep-wide surveys via unsupervised machine learning}",
      journal = {\mnras},
         year = 2020,
        month = jan,
       volume = {491},
       number = {1},
        pages = {1408-1426},
          doi = {10.1093/mnras/stz3006},
archivePrefix = {arXiv},
       eprint = {1909.10537},
 primaryClass = {astro-ph.GA},
       adsurl = {https://ui.adsabs.harvard.edu/abs/2020MNRAS.491.1408M}
}

@ARTICLE{Ciprijanovic23,
       author = {{{\'C}iprijanovi{\'c}}, A. and {Lewis}, A. and {Pedro}, K. and {Madireddy}, S. and {Nord}, B. and {Perdue}, G.~N. and {Wild}, S.~M.},
        title = "{DeepAstroUDA: semi-supervised universal domain adaptation for cross-survey galaxy morphology classification and anomaly detection}",
      journal = {Machine Learning: Science and Technology},
         year = 2023,
        month = jun,
       volume = {4},
       number = {2},
          eid = {025013},
        pages = {025013},
          doi = {10.1088/2632-2153/acca5f},
archivePrefix = {arXiv},
       eprint = {2302.02005},
 primaryClass = {astro-ph.GA},
       adsurl = {https://ui.adsabs.harvard.edu/abs/2023MLS&T...4b5013C}
}

@ARTICLE{Siudek25,
       author = {{Siudek}, M. and {Mezcua}, M. and {Circosta}, C. and {Maraston}, C. and {Moustakas}, J. and {Zou}, H. and {Aguilar}, J. and {Ahlen}, S. and {Bianchi}, D. and {Brooks}, D. and {Claybaugh}, T. and {Dawson}, K.~S. and {de la Macorra}, A. and {Dey}, A. and {Doel}, P. and {Forero-Romero}, J.~E. and {Gazta{\~n}aga}, E. and {Gontcho A Gontcho}, S. and {Gutierrez}, G. and {Ishak}, M. and {Juneau}, S. and {Kirkby}, D. and {Kisner}, T. and {Kremin}, A. and {Lambert}, A. and {Landriau}, M. and {Le Guillou}, L. and {Meisner}, A. and {Miquel}, R. and {Prada}, F. and {P{\'e}rez-R{\`a}fols}, I. and {Rossi}, G. and {Sanchez}, E. and {Schlegel}, D. and {Schubnell}, M. and {Seo}, H. and {Sprayberry}, D. and {Tarl{\'e}}, G. and {Weaver}, B.~A.},
        title = "{Beyond traditional diagnostics: Identifying active galactic nuclei using spectral energy distribution fitting in DESI data}",
      journal = {\aap},
         year = 2025,
        month = aug,
       volume = {700},
          eid = {A209},
        pages = {A209},
          doi = {10.1051/0004-6361/202555463},
archivePrefix = {arXiv},
       eprint = {2506.09143},
 primaryClass = {astro-ph.GA},
       adsurl = {https://ui.adsabs.harvard.edu/abs/2025A&A...700A.209S}
}

@ARTICLE{LSST19,
       author = {{Ivezi{\'c}}, {\v{Z}}eljko and {Kahn}, Steven M. and {Tyson}, J. Anthony and {Abel}, Bob and {Acosta}, Emily and {Allsman}, Robyn and {Alonso}, David and {AlSayyad}, Yusra and {Anderson}, Scott F. and {Andrew}, John and {Angel}, James Roger P. and {Angeli}, George Z. and {Ansari}, Reza and {Antilogus}, Pierre and {Araujo}, Constanza and {Armstrong}, Robert and {Arndt}, Kirk T. and {Astier}, Pierre and {Aubourg}, {\'E}ric and {Auza}, Nicole and {Axelrod}, Tim S. and {Bard}, Deborah J. and {Barr}, Jeff D. and {Barrau}, Aurelian and {Bartlett}, James G. and {Bauer}, Amanda E. and {Bauman}, Brian J. and {Baumont}, Sylvain and {Bechtol}, Ellen and {Bechtol}, Keith and {Becker}, Andrew C. and {Becla}, Jacek and {Beldica}, Cristina and {Bellavia}, Steve and {Bianco}, Federica B. and {Biswas}, Rahul and {Blanc}, Guillaume and {Blazek}, Jonathan and {Blandford}, Roger D. and {Bloom}, Josh S. and {Bogart}, Joanne and {Bond}, Tim W. and {Booth}, Michael T. and {Borgland}, Anders W. and {Borne}, Kirk and {Bosch}, James F. and {Boutigny}, Dominique and {Brackett}, Craig A. and {Bradshaw}, Andrew and {Brandt}, William Nielsen and {Brown}, Michael E. and {Bullock}, James S. and {Burchat}, Patricia and {Burke}, David L. and {Cagnoli}, Gianpietro and {Calabrese}, Daniel and {Callahan}, Shawn and {Callen}, Alice L. and {Carlin}, Jeffrey L. and {Carlson}, Erin L. and {Chandrasekharan}, Srinivasan and {Charles-Emerson}, Glenaver and {Chesley}, Steve and {Cheu}, Elliott C. and {Chiang}, Hsin-Fang and {Chiang}, James and {Chirino}, Carol and {Chow}, Derek and {Ciardi}, David R. and {Claver}, Charles F. and {Cohen-Tanugi}, Johann and {Cockrum}, Joseph J. and {Coles}, Rebecca and {Connolly}, Andrew J. and {Cook}, Kem H. and {Cooray}, Asantha and {Covey}, Kevin R. and {Cribbs}, Chris and {Cui}, Wei and {Cutri}, Roc and {Daly}, Philip N. and {Daniel}, Scott F. and {Daruich}, Felipe and {Daubard}, Guillaume and {Daues}, Greg and {Dawson}, William and {Delgado}, Francisco and {Dellapenna}, Alfred and {de Peyster}, Robert and {de Val-Borro}, Miguel and {Digel}, Seth W. and {Doherty}, Peter and {Dubois}, Richard and {Dubois-Felsmann}, Gregory P. and {Durech}, Josef and {Economou}, Frossie and {Eifler}, Tim and {Eracleous}, Michael and {Emmons}, Benjamin L. and {Fausti Neto}, Angelo and {Ferguson}, Henry and {Figueroa}, Enrique and {Fisher-Levine}, Merlin and {Focke}, Warren and {Foss}, Michael D. and {Frank}, James and {Freemon}, Michael D. and {Gangler}, Emmanuel and {Gawiser}, Eric and {Geary}, John C. and {Gee}, Perry and {Geha}, Marla and {Gessner}, Charles J.~B. and {Gibson}, Robert R. and {Gilmore}, D. Kirk and {Glanzman}, Thomas and {Glick}, William and {Goldina}, Tatiana and {Goldstein}, Daniel A. and {Goodenow}, Iain and {Graham}, Melissa L. and {Gressler}, William J. and {Gris}, Philippe and {Guy}, Leanne P. and {Guyonnet}, Augustin and {Haller}, Gunther and {Harris}, Ron and {Hascall}, Patrick A. and {Haupt}, Justine and {Hernandez}, Fabio and {Herrmann}, Sven and {Hileman}, Edward and {Hoblitt}, Joshua and {Hodgson}, John A. and {Hogan}, Craig and {Howard}, James D. and {Huang}, Dajun and {Huffer}, Michael E. and {Ingraham}, Patrick and {Innes}, Walter R. and {Jacoby}, Suzanne H. and {Jain}, Bhuvnesh and {Jammes}, Fabrice and {Jee}, M. James and {Jenness}, Tim and {Jernigan}, Garrett and {Jevremovi{\'c}}, Darko and {Johns}, Kenneth and {Johnson}, Anthony S. and {Johnson}, Margaret W.~G. and {Jones}, R. Lynne and {Juramy-Gilles}, Claire and {Juri{\'c}}, Mario and {Kalirai}, Jason S. and {Kallivayalil}, Nitya J. and {Kalmbach}, Bryce and {Kantor}, Jeffrey P. and {Karst}, Pierre and {Kasliwal}, Mansi M. and {Kelly}, Heather and {Kessler}, Richard and {Kinnison}, Veronica and {Kirkby}, David and {Knox}, Lloyd and {Kotov}, Ivan V. and {Krabbendam}, Victor L. and {Krughoff}, K. Simon and {Kub{\'a}nek}, Petr and {Kuczewski}, John and {Kulkarni}, Shri and {Ku}, John and {Kurita}, Nadine R. and {Lage}, Craig S. and {Lambert}, Ron and {Lange}, Travis and {Langton}, J. Brian and {Le Guillou}, Laurent and {Levine}, Deborah and {Liang}, Ming and {Lim}, Kian-Tat and {Lintott}, Chris J. and {Long}, Kevin E. and {Lopez}, Margaux and {Lotz}, Paul J. and {Lupton}, Robert H. and {Lust}, Nate B. and {MacArthur}, Lauren A. and {Mahabal}, Ashish and {Mandelbaum}, Rachel and {Markiewicz}, Thomas W. and {Marsh}, Darren S. and {Marshall}, Philip J. and {Marshall}, Stuart and {May}, Morgan and {McKercher}, Robert and {McQueen}, Michelle and {Meyers}, Joshua and {Migliore}, Myriam and {Miller}, Michelle and {Mills}, David J.},
        title = "{LSST: From Science Drivers to Reference Design and Anticipated Data Products}",
      journal = {\apj},
         year = 2019,
        month = mar,
       volume = {873},
       number = {2},
          eid = {111},
        pages = {111},
          doi = {10.3847/1538-4357/ab042c},
archivePrefix = {arXiv},
       eprint = {0805.2366},
 primaryClass = {astro-ph},
       adsurl = {https://ui.adsabs.harvard.edu/abs/2019ApJ...873..111I}
}

@ARTICLE{4MOST19,
       author = {{de Jong}, Roelof S.},
        title = "{A spectroscopy facility for many}",
      journal = {Nature Astronomy},
         year = 2019,
        month = jun,
       volume = {3},
        pages = {574-574},
          doi = {10.1038/s41550-019-0808-x},
       adsurl = {https://ui.adsabs.harvard.edu/abs/2019NatAs...3..574D}
}

@ARTICLE{Shao22,
       author = {{Shao}, Helen and {Villaescusa-Navarro}, Francisco and {Genel}, Shy and {Spergel}, David N. and {Angl{\'e}s-Alc{\'a}zar}, Daniel and {Hernquist}, Lars and {Dav{\'e}}, Romeel and {Narayanan}, Desika and {Contardo}, Gabriella and {Vogelsberger}, Mark},
        title = "{Finding Universal Relations in Subhalo Properties with Artificial Intelligence}",
      journal = {\apj},
         year = 2022,
        month = mar,
       volume = {927},
       number = {1},
          eid = {85},
        pages = {85},
          doi = {10.3847/1538-4357/ac4d30},
archivePrefix = {arXiv},
       eprint = {2109.04484},
 primaryClass = {astro-ph.CO},
       adsurl = {https://ui.adsabs.harvard.edu/abs/2022ApJ...927...85S}
}

@ARTICLE{Ofman22,
       author = {{Ofman}, Leon and {Averbuch}, Amir and {Shliselberg}, Adi and {Benaun}, Idan and {Segev}, David and {Rissman}, Aron},
        title = "{Automated identification of transiting exoplanet candidates in NASA Transiting Exoplanets Survey Satellite (TESS) data with machine learning methods}",
      journal = {\na},
         year = 2022,
        month = feb,
       volume = {91},
          eid = {101693},
        pages = {101693},
          doi = {10.1016/j.newast.2021.101693},
archivePrefix = {arXiv},
       eprint = {2102.10326},
 primaryClass = {astro-ph.EP},
       adsurl = {https://ui.adsabs.harvard.edu/abs/2022NewA...9101693O}
}

@ARTICLE{Baron19,
       author = {{Baron}, Dalya},
        title = "{Machine Learning in Astronomy: a practical overview}",
      journal = {arXiv e-prints},
         year = 2019,
        month = apr,
          eid = {arXiv:1904.07248},
        pages = {arXiv:1904.07248},
          doi = {10.48550/arXiv.1904.07248},
archivePrefix = {arXiv},
       eprint = {1904.07248},
 primaryClass = {astro-ph.IM},
       adsurl = {https://ui.adsabs.harvard.edu/abs/2019arXiv190407248B}
}

@INPROCEEDINGS{VanderPlas12,
       author = {{VanderPlas}, J. and {Connolly}, A.~J. and {Ivezic}, Z. and {Gray}, A.},
        title = "{Introduction to astroML: Machine learning for astrophysics}",
    booktitle = {Proceedings of Conference on Intelligent Data Understanding (CIDU},
         year = 2012,
        month = oct,
        pages = {47-54},
          doi = {10.1109/CIDU.2012.6382200},
archivePrefix = {arXiv},
       eprint = {1411.5039},
 primaryClass = {astro-ph.IM},
       adsurl = {https://ui.adsabs.harvard.edu/abs/2012cidu.conf...47V}
}

@ARTICLE{Ball10,
       author = {{Ball}, Nicholas M. and {Brunner}, Robert J.},
        title = "{Data Mining and Machine Learning in Astronomy}",
      journal = {International Journal of Modern Physics D},
         year = 2010,
        month = jan,
       volume = {19},
       number = {7},
        pages = {1049-1106},
          doi = {10.1142/S0218271810017160},
archivePrefix = {arXiv},
       eprint = {0906.2173},
 primaryClass = {astro-ph.IM},
       adsurl = {https://ui.adsabs.harvard.edu/abs/2010IJMPD..19.1049B}
}

@ARTICLE{Abbott21,
       author = {{Abbott}, T.~M.~C. and {Adam{\'o}w}, M. and {Aguena}, M. and {Allam}, S. and {Amon}, A. and {Annis}, J. and {Avila}, S. and {Bacon}, D. and {Banerji}, M. and {Bechtol}, K. and {Becker}, M.~R. and {Bernstein}, G.~M. and {Bertin}, E. and {Bhargava}, S. and {Bridle}, S.~L. and {Brooks}, D. and {Burke}, D.~L. and {Carnero Rosell}, A. and {Carrasco Kind}, M. and {Carretero}, J. and {Castander}, F.~J. and {Cawthon}, R. and {Chang}, C. and {Choi}, A. and {Conselice}, C. and {Costanzi}, M. and {Crocce}, M. and {da Costa}, L.~N. and {Davis}, T.~M. and {De Vicente}, J. and {DeRose}, J. and {Desai}, S. and {Diehl}, H.~T. and {Dietrich}, J.~P. and {Drlica-Wagner}, A. and {Eckert}, K. and {Elvin-Poole}, J. and {Everett}, S. and {Evrard}, A.~E. and {Ferrero}, I. and {Fert{\'e}}, A. and {Flaugher}, B. and {Fosalba}, P. and {Friedel}, D. and {Frieman}, J. and {Garc{\'\i}a-Bellido}, J. and {Gaztanaga}, E. and {Gelman}, L. and {Gerdes}, D.~W. and {Giannantonio}, T. and {Gill}, M.~S.~S. and {Gruen}, D. and {Gruendl}, R.~A. and {Gschwend}, J. and {Gutierrez}, G. and {Hartley}, W.~G. and {Hinton}, S.~R. and {Hollowood}, D.~L. and {Honscheid}, K. and {Huterer}, D. and {James}, D.~J. and {Jeltema}, T. and {Johnson}, M.~D. and {Kent}, S. and {Kron}, R. and {Kuehn}, K. and {Kuropatkin}, N. and {Lahav}, O. and {Li}, T.~S. and {Lidman}, C. and {Lin}, H. and {MacCrann}, N. and {Maia}, M.~A.~G. and {Manning}, T.~A. and {Maloney}, J.~D. and {March}, M. and {Marshall}, J.~L. and {Martini}, P. and {Melchior}, P. and {Menanteau}, F. and {Miquel}, R. and {Morgan}, R. and {Myles}, J. and {Neilsen}, E. and {Ogando}, R.~L.~C. and {Palmese}, A. and {Paz-Chinch{\'o}n}, F. and {Petravick}, D. and {Pieres}, A. and {Plazas}, A.~A. and {Pond}, C. and {Rodriguez-Monroy}, M. and {Romer}, A.~K. and {Roodman}, A. and {Rykoff}, E.~S. and {Sako}, M. and {Sanchez}, E. and {Santiago}, B. and {Scarpine}, V. and {Serrano}, S. and {Sevilla-Noarbe}, I. and {Smith}, J. Allyn and {Smith}, M. and {Soares-Santos}, M. and {Suchyta}, E. and {Swanson}, M.~E.~C. and {Tarle}, G. and {Thomas}, D. and {To}, C. and {Tremblay}, P.~E. and {Troxel}, M.~A. and {Tucker}, D.~L. and {Turner}, D.~J. and {Varga}, T.~N. and {Walker}, A.~R. and {Wechsler}, R.~H. and {Weller}, J. and {Wester}, W. and {Wilkinson}, R.~D. and {Yanny}, B. and {Zhang}, Y. and {Nikutta}, R. and {Fitzpatrick}, M. and {Jacques}, A. and {Scott}, A. and {Olsen}, K. and {Huang}, L. and {Herrera}, D. and {Juneau}, S. and {Nidever}, D. and {Weaver}, B.~A. and {Adean}, C. and {Correia}, V. and {de Freitas}, M. and {Freitas}, F.~N. and {Singulani}, C. and {Vila-Verde}, G. and {Linea Science Server}},
        title = "{The Dark Energy Survey Data Release 2}",
      journal = {\apjs},
         year = 2021,
        month = aug,
       volume = {255},
       number = {2},
          eid = {20},
        pages = {20},
          doi = {10.3847/1538-4365/ac00b3},
archivePrefix = {arXiv},
       eprint = {2101.05765},
 primaryClass = {astro-ph.IM},
       adsurl = {https://ui.adsabs.harvard.edu/abs/2021ApJS..255...20A}
}

@ARTICLE{Abbott18,
       author = {{Abbott}, T.~M.~C. and {Abdalla}, F.~B. and {Allam}, S. and {Amara}, A. and {Annis}, J. and {Asorey}, J. and {Avila}, S. and {Ballester}, O. and {Banerji}, M. and {Barkhouse}, W. and {Baruah}, L. and {Baumer}, M. and {Bechtol}, K. and {Becker}, M.~R. and {Benoit-L{\'e}vy}, A. and {Bernstein}, G.~M. and {Bertin}, E. and {Blazek}, J. and {Bocquet}, S. and {Brooks}, D. and {Brout}, D. and {Buckley-Geer}, E. and {Burke}, D.~L. and {Busti}, V. and {Campisano}, R. and {Cardiel-Sas}, L. and {Carnero Rosell}, A. and {Carrasco Kind}, M. and {Carretero}, J. and {Castander}, F.~J. and {Cawthon}, R. and {Chang}, C. and {Chen}, X. and {Conselice}, C. and {Costa}, G. and {Crocce}, M. and {Cunha}, C.~E. and {D'Andrea}, C.~B. and {da Costa}, L.~N. and {Das}, R. and {Daues}, G. and {Davis}, T.~M. and {Davis}, C. and {De Vicente}, J. and {DePoy}, D.~L. and {DeRose}, J. and {Desai}, S. and {Diehl}, H.~T. and {Dietrich}, J.~P. and {Dodelson}, S. and {Doel}, P. and {Drlica-Wagner}, A. and {Eifler}, T.~F. and {Elliott}, A.~E. and {Evrard}, A.~E. and {Farahi}, A. and {Fausti Neto}, A. and {Fernandez}, E. and {Finley}, D.~A. and {Flaugher}, B. and {Foley}, R.~J. and {Fosalba}, P. and {Friedel}, D.~N. and {Frieman}, J. and {Garc{\'\i}a-Bellido}, J. and {Gaztanaga}, E. and {Gerdes}, D.~W. and {Giannantonio}, T. and {Gill}, M.~S.~S. and {Glazebrook}, K. and {Goldstein}, D.~A. and {Gower}, M. and {Gruen}, D. and {Gruendl}, R.~A. and {Gschwend}, J. and {Gupta}, R.~R. and {Gutierrez}, G. and {Hamilton}, S. and {Hartley}, W.~G. and {Hinton}, S.~R. and {Hislop}, J.~M. and {Hollowood}, D. and {Honscheid}, K. and {Hoyle}, B. and {Huterer}, D. and {Jain}, B. and {James}, D.~J. and {Jeltema}, T. and {Johnson}, M.~W.~G. and {Johnson}, M.~D. and {Kacprzak}, T. and {Kent}, S. and {Khullar}, G. and {Klein}, M. and {Kovacs}, A. and {Koziol}, A.~M.~G. and {Krause}, E. and {Kremin}, A. and {Kron}, R. and {Kuehn}, K. and {Kuhlmann}, S. and {Kuropatkin}, N. and {Lahav}, O. and {Lasker}, J. and {Li}, T.~S. and {Li}, R.~T. and {Liddle}, A.~R. and {Lima}, M. and {Lin}, H. and {L{\'o}pez-Reyes}, P. and {MacCrann}, N. and {Maia}, M.~A.~G. and {Maloney}, J.~D. and {Manera}, M. and {March}, M. and {Marriner}, J. and {Marshall}, J.~L. and {Martini}, P. and {McClintock}, T. and {McKay}, T. and {McMahon}, R.~G. and {Melchior}, P. and {Menanteau}, F. and {Miller}, C.~J. and {Miquel}, R. and {Mohr}, J.~J. and {Morganson}, E. and {Mould}, J. and {Neilsen}, E. and {Nichol}, R.~C. and {Nogueira}, F. and {Nord}, B. and {Nugent}, P. and {Nunes}, L. and {Ogando}, R.~L.~C. and {Old}, L. and {Pace}, A.~B. and {Palmese}, A. and {Paz-Chinch{\'o}n}, F. and {Peiris}, H.~V. and {Percival}, W.~J. and {Petravick}, D. and {Plazas}, A.~A. and {Poh}, J. and {Pond}, C. and {Porredon}, A. and {Pujol}, A. and {Refregier}, A. and {Reil}, K. and {Ricker}, P.~M. and {Rollins}, R.~P. and {Romer}, A.~K. and {Roodman}, A. and {Rooney}, P. and {Ross}, A.~J. and {Rykoff}, E.~S. and {Sako}, M. and {Sanchez}, M.~L. and {Sanchez}, E. and {Santiago}, B. and {Saro}, A. and {Scarpine}, V. and {Scolnic}, D. and {Serrano}, S. and {Sevilla-Noarbe}, I. and {Sheldon}, E. and {Shipp}, N. and {Silveira}, M.~L. and {Smith}, M. and {Smith}, R.~C. and {Smith}, J.~A. and {Soares-Santos}, M. and {Sobreira}, F. and {Song}, J. and {Stebbins}, A. and {Suchyta}, E. and {Sullivan}, M. and {Swanson}, M.~E.~C. and {Tarle}, G. and {Thaler}, J. and {Thomas}, D. and {Thomas}, R.~C. and {Troxel}, M.~A. and {Tucker}, D.~L. and {Vikram}, V. and {Vivas}, A.~K. and {Walker}, A.~R. and {Wechsler}, R.~H. and {Weller}, J. and {Wester}, W. and {Wolf}, R.~C. and {Wu}, H. and {Yanny}, B. and {Zenteno}, A. and {Zhang}, Y. and {Zuntz}, J. and {DES Collaboration} and {Juneau}, S. and {Fitzpatrick}, M. and {Nikutta}, R.},
        title = "{The Dark Energy Survey: Data Release 1}",
      journal = {\apjs},
         year = 2018,
        month = dec,
       volume = {239},
       number = {2},
          eid = {18},
        pages = {18},
          doi = {10.3847/1538-4365/aae9f0},
archivePrefix = {arXiv},
       eprint = {1801.03181},
 primaryClass = {astro-ph.IM},
       adsurl = {https://ui.adsabs.harvard.edu/abs/2018ApJS..239...18A}
}

@ARTICLE{Beverage21,
       author = {{Beverage}, Aliza G. and {Kriek}, Mariska and {Conroy}, Charlie and {Bezanson}, Rachel and {Franx}, Marijn and {van der Wel}, Arjen},
        title = "{Elemental Abundances and Ages of z   0.7 Quiescent Galaxies on the Mass-Size Plane: Implication for Chemical Enrichment and Star Formation Quenching}",
      journal = {\apjl},
         year = 2021,
        month = aug,
       volume = {917},
       number = {1},
          eid = {L1},
        pages = {L1},
          doi = {10.3847/2041-8213/ac12cd},
archivePrefix = {arXiv},
       eprint = {2105.12750},
 primaryClass = {astro-ph.GA},
       adsurl = {https://ui.adsabs.harvard.edu/abs/2021ApJ...917L...1B}
}

@ARTICLE{DESI_I_2025,
       author = {{DESI Collaboration} and {Abdul-Karim}, M. and {Adame}, A.~G. and {Aguado}, D. and {Aguilar}, J. and {Ahlen}, S. and {Alam}, S. and {Aldering}, G. and {Alexander}, D.~M. and {Alfarsy}, R. and {Allen}, L. and {Allende Prieto}, C. and {Alves}, O. and {Anand}, A. and {Andrade}, U. and {Armengaud}, E. and {Avila}, S. and {Aviles}, A. and {Awan}, H. and {Bailey}, S. and {Baleato Lizancos}, A. and {Ballester}, O. and {Bault}, A. and {Bautista}, J. and {BenZvi}, S. and {Beraldo e Silva}, L. and {Bermejo-Climent}, J.~R. and {Beutler}, F. and {Bianchi}, D. and {Blake}, C. and {Blum}, R. and {Bolton}, A.~S. and {Bonici}, M. and {Brieden}, S. and {Brodzeller}, A. and {Brooks}, D. and {Buckley-Geer}, E. and {Burtin}, E. and {Canning}, R. and {Carnero Rosell}, A. and {Carr}, A. and {Carrilho}, P. and {Casas}, L. and {Castander}, F.~J. and {Cereskaite}, R. and {Cervantes-Cota}, J.~L. and {Chaussidon}, E. and {Chaves-Montero}, J. and {Chen}, S. and {Chen}, X. and {Claybaugh}, T. and {Cole}, S. and {Cooper}, A.~P. and {Cousinou}, M. -C. and {Cuceu}, A. and {Davis}, T.~M. and {Dawson}, K.~S. and {de Belsunce}, R. and {de la Cruz}, R. and {de la Macorra}, A. and {de Mattia}, A. and {Deiosso}, N. and {Della Costa}, J. and {Demina}, R. and {Demirbozan}, U. and {DeRose}, J. and {Dey}, A. and {Dey}, B. and {Ding}, J. and {Ding}, Z. and {Doel}, P. and {Douglass}, K. and {Dowicz}, M. and {Ebina}, H. and {Edelstein}, J. and {Eisenstein}, D.~J. and {Elbers}, W. and {Emas}, N. and {Escoffier}, S. and {Fagrelius}, P. and {Fan}, X. and {Fanning}, K. and {Fawcett}, V.~A. and {Fern\textbackslash'andez-Garc\textbackslash'ia}, E. and {Ferraro}, S. and {Findlay}, N. and {Font-Ribera}, A. and {Forero-Romero}, J.~E. and {Forero-S\textbackslash'anchez}, D. and {Frenk}, C.~S. and {G\textbackslash''ansicke}, B.~T. and {Galbany}, L. and {Garc\textbackslash'ia-Bellido}, J. and {Garcia-Quintero}, C. and {Garrison}, L.~H. and {Gazta\textbackslash\raisebox{-0.5ex}\textasciitildenaga}, E. and {Gil-Mar\textbackslash'in}, H. and {Gnedin}, O.~Y. and {Gontcho}, S. Gontcho A and {Gonzalez-Morales}, A.~X. and {Gonzalez-Perez}, V. and {Gordon}, C. and {Graur}, O. and {Green}, D. and {Gruen}, D. and {Gsponer}, R. and {Guandalin}, C. and {Gutierrez}, G. and {Guy}, J. and {Hahn}, C. and {Han}, J.~J. and {Han}, J. and {He}, S. and {Herrera-Alcantar}, H.~K. and {Honscheid}, K. and {Hou}, J. and {Howlett}, C. and {Huterer}, D. and {Ir\textbackslashv\{s\}i\textbackslashv\{c\}}, V. and {Ishak}, M. and {Jacques}, A. and {Jimenez}, J. and {Jing}, Y.~P. and {Joachimi}, B. and {Joudaki}, S. and {Joyce}, R. and {Jullo}, E. and {Juneau}, S. and {Kara\textbackslashc\{c\}ayl\{\textbackslashi\}}, N.~G. and {Karim}, T. and {Kehoe}, R. and {Kent}, S. and {Khederlarian}, A. and {Kirkby}, D. and {Kisner}, T. and {Kitaura}, F. -S. and {Kizhuprakkat}, N. and {Kong}, H. and {Koposov}, S.~E. and {Kremin}, A. and {Krolewski}, A. and {Lahav}, O. and {Lai}, Y. and {Lamman}, C. and {Lan}, T. -W. and {Landriau}, M. and {Lang}, D. and {Lange}, J.~U. and {Lasker}, J. and {Le Goff}, J.~M. and {Le Guillou}, L. and {Leauthaud}, A. and {Levi}, M.~E. and {Li}, S. and {Li}, T.~S. and {Lodha}, K. and {Lokken}, M. and {Luo}, Y. and {Magneville}, C. and {Manera}, M. and {Manser}, C.~J. and {Margala}, D. and {Martini}, P. and {Maus}, M. and {McCullough}, J. and {McDonald}, P. and {Medina}, G.~E. and {Medina-Varela}, L. and {Meisner}, A. and {Mena-Fern\textbackslash'andez}, J. and {Menegas}, A. and {Mezcua}, M. and {Miquel}, R. and {Montero-Camacho}, P. and {Moon}, J. and {Moustakas}, J. and {Mu\textbackslash\raisebox{-0.5ex}\textasciitildenoz-Guti\textbackslash'errez}, A. and {Mu\textbackslash\raisebox{-0.5ex}\textasciitildenoz-Santos}, D. and {Myers}, A.~D. and {Myles}, J. and {Nadathur}, S. and {Najita}, J. and {Napolitano}, L. and {Newman}, J.~A. and {Nikakhtar}, F. and {Nikutta}, R. and {Niz}, G. and {Noriega}, H.~E. and {Padmanabhan}, N. and {Paillas}, E. and {Palanque-Delabrouille}, N. and {Palmese}, A. and {Pan}, J. and {Pan}, Z. and {Parkinson}, D. and {Peacock}, J. and {Percival}, W.~J. and {P\textbackslash'erez-Fern\textbackslash'andez}, A. and {P\textbackslash'erez-R\textbackslash`afols}, I. and {Peterson}, P.},
        title = "{Data Release 1 of the Dark Energy Spectroscopic Instrument}",
      journal = {arXiv e-prints},
         year = 2025,
        month = mar,
          eid = {arXiv:2503.14745},
        pages = {arXiv:2503.14745},
          doi = {10.48550/arXiv.2503.14745},
archivePrefix = {arXiv},
       eprint = {2503.14745},
 primaryClass = {astro-ph.CO},
       adsurl = {https://ui.adsabs.harvard.edu/abs/2025arXiv250314745D}
}

@ARTICLE{Khramtsov21,
       author = {{Khramtsov}, V. and {Spiniello}, C. and {Agnello}, A. and {Sergeyev}, A.},
        title = "{VEXAS: VISTA EXtension to Auxiliary Surveys. Data Release 2: Machine-learning based classification of sources in the Southern Hemisphere}",
      journal = {\aap},
         year = 2021,
        month = jul,
       volume = {651},
          eid = {A69},
        pages = {A69},
          doi = {10.1051/0004-6361/202040131},
archivePrefix = {arXiv},
       eprint = {2103.09257},
 primaryClass = {astro-ph.GA},
       adsurl = {https://ui.adsabs.harvard.edu/abs/2021A&A...651A..69K}
}

@ARTICLE{breiman2001random,
  title={Random forests},
  author={Breiman, Leo},
  journal={Machine learning},
  volume={45},
  pages={5--32},
  year={2001},
  publisher={Springer},
  doi = {10.1023/A:1010933404324},
}

@ARTICLE{Scaramella22,
       author = {{Euclid Collaboration: Scaramella} and {Scaramella}, R. and {Amiaux}, J. and {Mellier}, Y. and {Burigana}, C. and {Carvalho}, C.~S. and {Cuillandre}, J. -C. and {Da Silva}, A. and {Derosa}, A. and {Dinis}, J. and {Maiorano}, E. and {Maris}, M. and {Tereno}, I. and {Laureijs}, R. and {Boenke}, T. and {Buenadicha}, G. and {Dupac}, X. and {Gaspar Venancio}, L.~M. and {G{\'o}mez-{\'A}lvarez}, P. and {Hoar}, J. and {Lorenzo Alvarez}, J. and {Racca}, G.~D. and {Saavedra-Criado}, G. and {Schwartz}, J. and {Vavrek}, R. and {Schirmer}, M. and {Aussel}, H. and {Azzollini}, R. and {Cardone}, V.~F. and {Cropper}, M. and {Ealet}, A. and {Garilli}, B. and {Gillard}, W. and {Granett}, B.~R. and {Guzzo}, L. and {Hoekstra}, H. and {Jahnke}, K. and {Kitching}, T. and {Maciaszek}, T. and {Meneghetti}, M. and {Miller}, L. and {Nakajima}, R. and {Niemi}, S.~M. and {Pasian}, F. and {Percival}, W.~J. and {Pottinger}, S. and {Sauvage}, M. and {Scodeggio}, M. and {Wachter}, S. and {Zacchei}, A. and {Aghanim}, N. and {Amara}, A. and {Auphan}, T. and {Auricchio}, N. and {Awan}, S. and {Balestra}, A. and {Bender}, R. and {Bodendorf}, C. and {Bonino}, D. and {Branchini}, E. and {Brau-Nogue}, S. and {Brescia}, M. and {Candini}, G.~P. and {Capobianco}, V. and {Carbone}, C. and {Carlberg}, R.~G. and {Carretero}, J. and {Casas}, R. and {Castander}, F.~J. and {Castellano}, M. and {Cavuoti}, S. and {Cimatti}, A. and {Cledassou}, R. and {Congedo}, G. and {Conselice}, C.~J. and {Conversi}, L. and {Copin}, Y. and {Corcione}, L. and {Costille}, A. and {Courbin}, F. and {Degaudenzi}, H. and {Douspis}, M. and {Dubath}, F. and {Duncan}, C.~A.~J. and {Dusini}, S. and {Farrens}, S. and {Ferriol}, S. and {Fosalba}, P. and {Fourmanoit}, N. and {Frailis}, M. and {Franceschi}, E. and {Franzetti}, P. and {Fumana}, M. and {Gillis}, B. and {Giocoli}, C. and {Grazian}, A. and {Grupp}, F. and {Haugan}, S.~V.~H. and {Holmes}, W. and {Hormuth}, F. and {Hudelot}, P. and {Kermiche}, S. and {Kiessling}, A. and {Kilbinger}, M. and {Kohley}, R. and {Kubik}, B. and {K{\"u}mmel}, M. and {Kunz}, M. and {Kurki-Suonio}, H. and {Lahav}, O. and {Ligori}, S. and {Lilje}, P.~B. and {Lloro}, I. and {Mansutti}, O. and {Marggraf}, O. and {Markovic}, K. and {Marulli}, F. and {Massey}, R. and {Maurogordato}, S. and {Melchior}, M. and {Merlin}, E. and {Meylan}, G. and {Mohr}, J.~J. and {Moresco}, M. and {Morin}, B. and {Moscardini}, L. and {Munari}, E. and {Nichol}, R.~C. and {Padilla}, C. and {Paltani}, S. and {Peacock}, J. and {Pedersen}, K. and {Pettorino}, V. and {Pires}, S. and {Poncet}, M. and {Popa}, L. and {Pozzetti}, L. and {Raison}, F. and {Rebolo}, R. and {Rhodes}, J. and {Rix}, H. -W. and {Roncarelli}, M. and {Rossetti}, E. and {Saglia}, R. and {Schneider}, P. and {Schrabback}, T. and {Secroun}, A. and {Seidel}, G. and {Serrano}, S. and {Sirignano}, C. and {Sirri}, G. and {Skottfelt}, J. and {Stanco}, L. and {Starck}, J.~L. and {Tallada-Cresp{\'\i}}, P. and {Tavagnacco}, D. and {Taylor}, A.~N. and {Teplitz}, H.~I. and {Toledo-Moreo}, R. and {Torradeflot}, F. and {Trifoglio}, M. and {Valentijn}, E.~A. and {Valenziano}, L. and {Verdoes Kleijn}, G.~A. and {Wang}, Y. and {Welikala}, N. and {Weller}, J. and {Wetzstein}, M. and {Zamorani}, G. and {Zoubian}, J. and {Andreon}, S. and {Baldi}, M. and {Bardelli}, S. and {Boucaud}, A. and {Camera}, S. and {Di Ferdinando}, D. and {Fabbian}, G. and {Farinelli}, R. and {Galeotta}, S. and {Graci{\'a}-Carpio}, J. and {Maino}, D. and {Medinaceli}, E. and {Mei}, S. and {Neissner}, C. and {Polenta}, G. and {Renzi}, A. and {Romelli}, E. and {Rosset}, C. and {Sureau}, F. and {Tenti}, M. and {Vassallo}, T. and {Zucca}, E. and {Baccigalupi}, C. and {Balaguera-Antol{\'\i}nez}, A. and {Battaglia}, P. and {Biviano}, A. and {Borgani}, S. and {Bozzo}, E. and {Cabanac}, R. and {Cappi}, A.},
        title = "{Euclid preparation. I. The Euclid Wide Survey}",
      journal = {\aap},
         year = 2022,
        month = jun,
       volume = {662},
          eid = {A112},
        pages = {A112},
          doi = {10.1051/0004-6361/202141938},
archivePrefix = {arXiv},
       eprint = {2108.01201},
 primaryClass = {astro-ph.CO},
       adsurl = {https://ui.adsabs.harvard.edu/abs/2022A&A...662A.112E}
}

@ARTICLE{Conselice22,
       author = {{Conselice}, Christopher J. and {Mundy}, Carl J. and {Ferreira}, Leonardo and {Duncan}, Kenneth},
        title = "{A Direct Measurement of Galaxy Major and Minor Merger Rates and Stellar Mass Accretion Histories at Z < 3 Using Galaxy Pairs in the REFINE Survey}",
      journal = {\apj},
         year = 2022,
        month = dec,
       volume = {940},
       number = {2},
          eid = {168},
        pages = {168},
          doi = {10.3847/1538-4357/ac9b1a},
archivePrefix = {arXiv},
       eprint = {2207.03984},
 primaryClass = {astro-ph.GA},
       adsurl = {https://ui.adsabs.harvard.edu/abs/2022ApJ...940..168C}
}

@ARTICLE{Ownsworth14,
       author = {{Ownsworth}, Jamie R. and {Conselice}, Christopher J. and {Mortlock}, Alice and {Hartley}, William G. and {Almaini}, Omar and {Duncan}, Ken and {Mundy}, Carl J.},
        title = "{Minor versus major mergers: the stellar mass growth of massive galaxies from z = 3 using number density selection techniques}",
      journal = {\mnras},
         year = 2014,
        month = dec,
       volume = {445},
       number = {3},
        pages = {2198-2213},
          doi = {10.1093/mnras/stu1802},
archivePrefix = {arXiv},
       eprint = {1409.1582},
 primaryClass = {astro-ph.GA},
       adsurl = {https://ui.adsabs.harvard.edu/abs/2014MNRAS.445.2198O}
}

@INPROCEEDINGS{Worthey+99,
       author = {{Worthey}, G.},
        title = "{The Age-Metallicity Degeneracy}",
    booktitle = {Spectrophotometric Dating of Stars and Galaxies},
         year = 1999,
       editor = {{Hubeny}, Ivan and {Heap}, Sally and {Cornett}, Robert},
       series = {Astronomical Society of the Pacific Conference Series},
       volume = {192},
        month = jan,
        pages = {283},
       adsurl = {https://ui.adsabs.harvard.edu/abs/1999ASPC..192..283W}
}

@ARTICLE{DAgo23,
       author = {{D'Ago}, G. and {Spiniello}, C. and {Coccato}, L. and {Tortora}, C. and {La Barbera}, F. and {Arnaboldi}, M. and {Bevacqua}, D. and {Ferr{\'e}-Mateu}, A. and {Gallazzi}, A. and {Hartke}, J. and {Hunt}, L.~K. and {Mart{\'\i}n-Navarro}, I. and {Napolitano}, N.~R. and {Pulsoni}, C. and {Radovich}, M. and {Saracco}, P. and {Scognamiglio}, D. and {Zibetti}, S.},
        title = "{INSPIRE: INvestigating Stellar Population In RElics. III. Second data release (DR2): testing the systematics on the stellar velocity dispersion}",
      journal = {\aap},
         year = 2023,
        month = apr,
       volume = {672},
          eid = {A17},
        pages = {A17, INSPIRE DR2},
          doi = {10.1051/0004-6361/202245542},
archivePrefix = {arXiv},
       eprint = {2302.05453},
 primaryClass = {astro-ph.GA},
       adsurl = {https://ui.adsabs.harvard.edu/abs/2023A&A...672A..17D}
}

@ARTICLE{Cappellari17,
   author = {{Cappellari}, M.},
    title = "{Improving the full spectrum fitting method: accurate convolution with Gauss-Hermite functions}",
  journal = {\mnras},
archivePrefix = "arXiv",
   eprint = {1607.08538},
     year = 2017,
    month = apr,
   volume = 466,
    pages = {798-811},
      doi = {10.1093/mnras/stw3020},
   adsurl = {http://adsabs.harvard.edu/abs/2017MNRAS.466..798C}
}

\bsp	
\label{lastpage}
\end{document}